\documentclass[structabstract]{aa} 
\usepackage{graphicx}
\usepackage{longtable}		
\usepackage[a4paper=true,hypertex]{hyperref}
\usepackage{txfonts}
\usepackage{natbib}
\usepackage{epsfig}
\usepackage{longtable}
\usepackage{times}
\usepackage[english]{babel}
\usepackage{url}
\bibpunct{(}{)}{;}{a}{}{,}

\newcommand {\nh} {$N_{\rm H}$ }
\newcommand {\ergsec}{\hbox{erg s$^{-1}$}}

\newcommand {\cmmdue} {cm$^{-2}$}
\newcommand {\hcm} {\hbox {\ifmmode $ atom cm$^{-2}\else atom cm$^{-2}$\fi}}
\begin{document}

\title{A complete X-ray spectral coverage of the 2010 May-June outbursts of Circinus X-1}
\authorrunning{A. D'A\`i et al.}
\titlerunning{Cir X-1 spectral evolution in outburst}
\author{A. D'A\`i$^{1}$, E. Bozzo$^{2}$, A. Papitto$^{3,4}$, R. Iaria$^{1}$, T. Di Salvo$^{1}$, L. Burderi$^{3}$, A. Riggio$^{5}$, 
E. Egron$^{3}$, N.R. Robba$^{1}$}
\institute{Dipartimento di Fisica, Universit\`a di Palermo, Via Archirafi 36, I-90123, Palermo, Italy\\ \email{antonino.dai@unipa.it} 
\and
ISDC Science Data Center for Astrophysics of the University of Geneva, chemin d\'Ecogia, 16, 1290, Versoix, Switzerland
\and
Dipartimento di Fisica, Universit\`a  degli Studi di Cagliari, SP Monserrato-Sestu, KM 0.7, 09042 Monserrato, Italy
\and
Institut de Ci\`encies de l'Espai (IEEC-CSIC), Facultat de Ci\`ncies, Campus UAB, Torre C5, Pares, 2da Planta, 08193-Bellaterra, Spain
\and
INAF – Osservatorio Astronomico di Cagliari, Poggio dei Pini, Strada 54, 09012 Capoterra, Italy
}

\date{}

\abstract{}{}{}{}{} 

\abstract{Circinus X-1  is a neutron-star-accreting X-ray  binary in a
  wide (P$_{\rm orb}$ = 16.6  d), eccentric orbit.  After two years of
  relatively low X-ray luminosity, in  May 2010 Circinus X-1 went into
  outburst, reaching  0.4 Crab flux.   This outburst lasted  for about
  two orbital cycles  and was followed by another  shorter and fainter
  outburst in June.}
  % aims heading (mandatory) 
{We focus here on the broadband X-ray spectral evolution of the source
  as it spans  about three order of magnitudes in  flux. We attempt to
  relate  luminosity, spectral  shape, local  absorption,  and orbital
  phase.} 
 {We  use multiple Rossi-XTE/PCA (3.0--25  keV) and Swift/XRT
  (1.0--9.0  keV)   observations  and   a  20  ks   long  Chandra/HETGS
  observation  (1.0--9.0 keV),  to comprehensively  track  the spectral
  evolution of  the source during  all the outbursting  phases.  These
  observations were  taken every two/three  days and cover  about four
  orbital cycles.  The PCA data  mostly cover the major  outburst, the
  XRT data monitor  the declining phase of the  major outburst and all
  the  phases of  the  minor  outburst, and  Chandra  data provide  an
  essential snapshot of the end of this overall outbursting phase.}
 % results  heading  (mandatory) 
  {The  X-ray spectrum can  be satisfactorily  described by  a thermal
    Comptonization model with variable neutral local absorption in all
    phases  of   the  outburst.    No  other  additive   component  is
    statistically  required. The  first  outburst decays linearly, with an \textit{ankle}
    in the light curve as the flux  decreases below  $\sim$\,5
    $\times$ 10$^{-10}$ erg cm$^{-2}$  s$^{-1}$. At the same time, the
    source shows  a clear spectral state transition  from an optically
    thick to an optically thin state. While the characteristics of the
    first,   bright,   outburst   can   be  interpreted   within   the
    disk-instability  scenario, the  following, minor,  outburst shows
    peculiarities that cannot be easily reconciled in this framework.}
  {} 
   \maketitle
%

%%%%%%%%%%%%%%%%%%%%%%
%%%%%%%%%%%%%%%%%%%%%%
%%%%%%%%%%%%%%%%%%%%%%
%%%%%%%%%%%%%%%%%%%%%%
%%%%%%%%%%%%%%%%%%%%%%
%%%%%%%%%%%%%%%%%%%%%% 

\section{Introduction}

Circinus  X-1 (Cir  X-1)  is one  of  the most  peculiar, though  most
well-studied accreting X-ray binary  (XRB) of the Galaxy. Almost daily
monitoring of this  source, with the Rossi-XTE All  Sky Monitor (ASM),
starting  from the  beginning of  1996 has  provided  an uninterrupted
insight into  the X-ray history of  this source in the  past 15 years;
from 1996 up to 2000, it was one of the brightest sources of the X-ray
sky (at  $\sim$\,3 Crab level), but  from 2000 to  2006 its luminosity
steadily decreased,  reaching in 2007  a new state characterised  by a
persistent   mCrab  flux   level,   with  sporadic   several-week-long
outbursts.   Chandra  observations during  a  faint  X-ray state  have
revealed a complex, parsec-scale, X-ray emitting structure, consisting
of spatially resolved X-ray  arcsec-long jets, embedded in an inflated
arcmin  radio lobe  structure, possibly  powered by  the  jet activity
\citep{stewart93,         tudose06,        heinz07,        soleri09a}.
\citet{miller-jones11}   resolved   the  jet   emission   up  to   the
milliarcsecond-scale,   distinguishing  the   symmetry   of  the   jet
structure, which has  an extent of the order of  $\sim$\,150 AU, and a
variability consistent with a mildly relativistic jet-speed.

The nature  of the  compact object, neutron  star (NS),  or black-hole
(BH) has  long been  debated, because Cir  X-1 shared  spectral timing
properties amid  the BH and NS  class.  The resumption in  May 2010 of
bursting  activity  \citep{papitto_atel_10,  linares10},  definitively
confirmed  the  initial  discovery  of  type-I X-ray  bursts  in  1986
\citep{tennant86}, and its association with the NS class.

A periodicity  of $\sim$\,16.6  days observed in  different wavebands,
from radio  to X-rays,  is associated with  the orbital period  of the
system.  The characteristic timescale  for the orbital evolution ($P/2
\dot{P}$) is very short, $\sim$\,1400  yr, suggesting that Cir X-1 may
be a very young system  \citep{saz03}.  The NS magnetic field might be
in-between   those  of   typical  old   low-mass  X-ray   binaries  (B
$\sim$\,10$^8$--10$^9$     G)     and     high-mass     systems     (B
$\sim$\,10$^{11}$--10$^{12}$  G), thus  permitting  the occurrence  of
X-ray bursts at lower accretion rates.

The X-ray  emission from Cir X-1  is also very peculiar.   It has been
found  that Cir  X-1 shows  strong variability  caused by  its rapidly
varying mass accretion rate and  the global physical conditions of the
local environment.   At high  luminosity (probably near  the Eddington
limit),  the   X-ray  broadband  (0.1--200  keV)   spectrum  could  be
satisfactorily described in terms  of Comptonized emission, where soft
photons of 0.4--0.5 keV are upscattered in a cloud of moderate optical
depth  (14   $<\tau<$  20)  and  0.8--1.7   keV  electron  temperature
\citep{iaria01b,  iaria01a,  iaria02}.   The  inferred radius  of  the
Comptonizing  plasma (90  km $<R_w<$  160 km),  was found  to  be much
greater than the classical NS dimensions, suggesting that we observe a
truncated  accretion disk.   Other components  have  occasionally been
detected,  in  both  the   hard  X-ray  domain  \citep[above  15  keV,
][]{iaria01b}  and   at  very   soft  energies  \citep[below   1  keV,
][]{iaria02}.  The  former was interpreted as a  possible signature of
non-thermal  Comptonization, whereas  the latter  was more  elusive to
constrain spectrally, owing to the possible presence of a warm ionised
absorber that  would produce multiple  absorption edges in  this range
\citep{iaria05}.   In  addition to  this,  often  near the  periastron
passage, a  variable cold absorber,  with column density values  up to
$10^{24}$  cm$^{-2}$,  can  partially  occult  the  primary  emission,
causing multiple dips in the light curve \citep{brandt96,shirey99}.

High-resolution  Chandra  observations  revealed P-Cygni  profiles  of
highly  ionised  elements,  which  are  indicative  of  a  radiatively
supported accretion-disk  wind \citep{brandt00,schulz02} seen  at high
inclination  angle.  Subsequent  Chandra observations,  performed when
the source accreted at substantially lower accretion rates, revealed a
complex pattern of absorption features of highly ionised elements, the
latter  of which  emerged as  the source  entered  a several-hour-long
period of flaring activity  \citep{dai07a}.  The discovery of a highly
ionised  absorber, strongly  dependent on  the accretion  rate  of the
source, favoured  again a close  to edge-on geometry,  where optically
thick  disk material is  radiatively uplifted  from the  disk surface.
However,   the   measured    jet   inclination   angle   \citep[$\sim$
5$^{\circ}$][]{fender00} and  absence of  a clear regularity  in dips,
which  is  typical  of  dipping  X-ray systems,  place  in  doubt  the
suitability of this scenario.

The companion star has not yet been clearly resolved. \citet{jonker07}
suggested that there is a possible giant optical counterpart, based on
phase-resolved  \textit{I}-band optical spectroscopic  and photometric
observations.   These  spectra display  evidence  of broad  absorption
Paschen lines  (but in emission during the  phase-zero passage), which
is consistent with a B5-A0  spectral type.  A 1.4 M$_{\odot}$ NS would
constrain the inclination angle and the companion mass to be $\gtrsim$
13$^{\circ}$.7 and $\lesssim$ 10 M$_{\odot}$, respectively.

The source distance is  poorly constrained. Early suggestions based on
radio  HI  absorption  features  suggested  a lower  limit  of  8  kpc
\citep{goss77},  while an estimate  based on  the energetics  of X-ray
bursts indicated a range between  7.8 kpc and 10 kpc \citep{jonker04}.
Distance based on the equivalent  X-ray hydrogen column estimate and a
re-analysis  of  the  HI   velocity  drifts  of  the  companion  start
\citep{mignani02} led  to a different estimate  of 4.1\,$\pm$\,0.3 kpc
\citep{iaria05}.  To  be consistent with the  most recent literature's
reported  luminosities,  we  assume,  hereafter,  a  distance  of  7.8
kpc. Luminosities for a distance of 4.1 kpc should then be scaled by a
factor of $\sim$\,0.3.

%%%%%%%%%%%%%%%%%%%%%%%%%%%%%%%%%%%%%%%%%%%%%%
%%%%%%%%%%%%%%%%%%%%%%%%%%%%%%%%%%%%%%%%%%%%%%

\section{Observations, light curves, and data reduction}

\begin{figure}[ht!]
   \centering
   \includegraphics[angle=-90, width=\columnwidth]{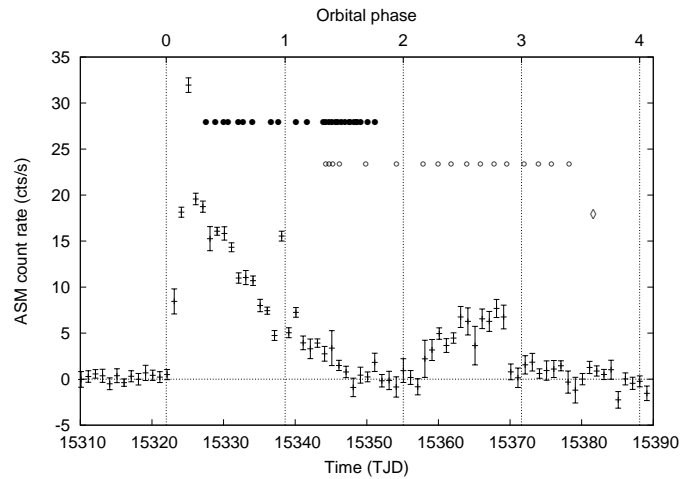}
   \caption{Daily-averaged ASM light curve (2--12  keV range) of  Cir X-1 between
     2010 April 19 and 2010 July 13. Markers for the RXTE (filled circles), 
      Swift (open circles), and Chandra (diamond) observations are superimposed. 
      Phase zero orbital passages are marked by grid lines.}
    \label{asm_countrate}
    \end{figure}

Starting from June 2008, the source entered a prolonged state of X-ray
faintness at  the few mCrab level,  which was interrupted  by a bright
outburst,  peaking at  0.4 Crab  flux, on  2010 May  9.   The outburst
started at $\sim$\,0.1 phase after the TJD 15322.04 phase-zero passage
according  to the  ephemeris  of \citet{nicolson07}\footnote{We  note,
  here, that calculating the  phase passage according to the ephemeris
  based on  the occurrence of the X-ray  dips \citep{clarkson04} would
  result  in  very  different  phase  value  (a  phase  difference  of
  $\sim$\,0.5) for  the outburst start  time.  Starting from  2004, we
  verified that  Nicolson's radio-based ephemeris are  able to predict
  the  start of  most of  the  X-ray outbursts,  while the  Clarkson's
  ephemeris shows  a progressive and  clear lagging of the  zero phase
  with  respect  to  the  peaks  of the  outbursts.  Because  of  this
  discrepancy,  we   refer,  hereafter,  only   to  the  orbital-phase
  calculations  based on  the radio  ephemeris.}, with  a  rising time
shorter than two days.  As the luminosity decreased, after two orbital
cycles, another outburst, peaking at $\sim$\,0.1 Crab, was detected on
2010    June   12.    We    present   in    the    upper   panel    of
Fig.\ref{asm_countrate}, the  RXTE ASM light  curve of Cir  X-1 during
the whole April--July 2010 period, with superimposed the visits of the
RXTE,  Swift-XRT, and  Chandra follow-ups.  Dotted-lines  indicate the
phase zero passages according to the ephemeris of \citet{nicolson07}.

\subsection{The RXTE monitoring campaign} \label{rxte_monitoring}

We studied  RXTE observations of Cir X-1  from 2010 May 11  to June 3.
We extracted and reduced data  only from the Proportional Counter Unit
(PCU)  2,  since this  counter  had  the  highest PCU  exposure  time,
applying standard filtering criteria.  We created additional good time
intervals to exclude all type-I X-ray bursts eventually present in the
data.   The burst  properties were  studied in  \citet{linares10}.  We
used  the   3.0--25.0  keV  energy   range  (3.0--20  keV,   when  the
signal-to-noise  ratio was  too poor  above 20  keV) and  a systematic
error of 1\%  was added in quadrature to the  best-fit model.  We also
cross-checked our  spectral results using only  the top-layer spectra,
finding  self-consistent  results.   We  applied  the  \textit{bright}
background  model   to  extract  the  background   spectra  and  light
curves. We verified  that when the source becomes faint  at the end of
the  outburst, the choice  between \textit{bright}  and \textit{faint}
background models does  not alter the conclusions of  our analysis.  A
log of all the observations (hereafter, we refer to the single pointed
RXTE  observation  using  the  last  four digits  of  its  observation
identifier,  or   ObsId)  used  in   the  analysis  is   presented  in
Table~\ref{table_rxte_obs}.

\begin{figure}[ht!]
  \centering
\includegraphics[angle=-90,width=\columnwidth]{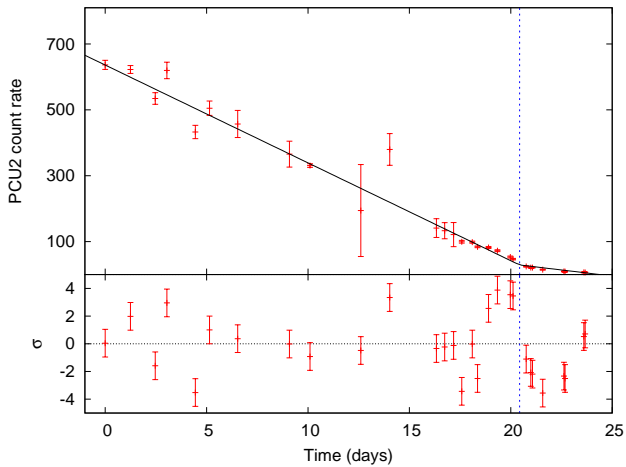}
\caption{Best-fit  model for the  RXTE light  curve (upper  panel) and
  residuals in units of standard deviations (lower panel).  Each point
  represents the average net (2--60 keV) light-curve count-rate in one
  ObsId. Time 0 corresponds to TJD 15327.0261.}
    \label{rxte_lc_fit}%
    \end{figure}

We  present   in  Fig.\ref{rxte_lc_fit}  the  2--60   keV  RXTE  light
curve. Each point represents the average, background-subtracted, count
rate  during   a  pointed  RXTE  observation.   The   light  curve  is
well-described by a combination of two linear decays; the first linear
covers  the first 20.5  days from  the start  of the  RXTE monitoring,
whereas a change  in the slope of the linear fit  covers the last five
days of the RXTE monitoring  (alternatively, an exponential fit with a
1.35 d decay-time also gives a good fit to this last set of data).  We
present in  the lower  panel the residuals  in units of  $\sigma$ with
respect to this phenomenological model of two linear decays, where the
two lines join at day 20.433  (TJD 15347.4591) and we also discern the
mid-time  between the  two observations  where a  clear change  in the
spectral  shape  is  also observed  (see  Sect.\ref{rxte_monitoring}).
Since the slope decay flattens  after the transition, we refer to this
feature as an \textit{ankle}.

\begin{figure}[ht!]
   \centering
   \includegraphics[angle=-90, width=8.5cm]{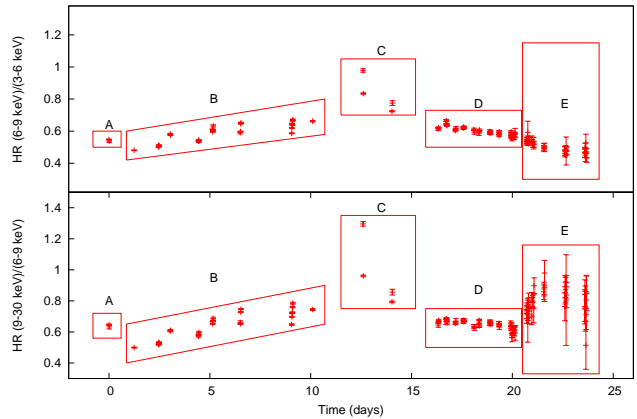}
   \caption{Hardness  ratios of the  May 2010  Cir X-1  outburst. Each
     point represents the ratio of the  average count rates in a 256 s
     time interval for the energy bands reported in $y$-labels.}
    \label{rxte_hid}%
    \end{figure}

To characterise the  spectral evolution of the source  during the RXTE
campaign  in  a  model-independent  way,  we  produced  two  different
hardness ratios  (HR), for the energy  bands: 3--6 keV,  6--9 keV, and
9--30 keV.  In  the upper panel of Fig.\ref{rxte_hid},  we show the HR
of  (6--9) keV/(3--6) keV  count rates  (soft HR),  and, in  the lower
panel, the  HR of  (9--30) keV/(6--9) keV  count rates (hard  HR). The
panels show different trends  as the outburst evolves.  We distinguish
five different regions: the region A covers the first RXTE observation
at the highest  count rate/luminosity; region B covers  about ten days
as the spectrum progressively becomes  harder, both in the soft and in
the  hard  ratios;  region   C  is  characterised  by  variable  local
absorption at the periastron passage; region D and E show a softening,
in the  soft ratio; and region E  shows a clear hardening  only in the
hard ratio.  We give a  model-dependent description of these trends in
Sect.~\ref{sec_analisi}.

\subsection{The Swift monitoring campaign}  \label{swift_monitoring}

Swift  monitored Cir  X-1 for  two orbital  cycles, from  2010  May 27
19:47:15 UT to 2010 June  30 21:32:57 UT.  Swift regularly visited the
source every two/five days. According to the Swift/XRT count rate, the
XRT data-collecting  mode switched  from the photon-counting  mode (PC
mode), for low  count rates, to the window-timing  mode (WT mode), for
high count  rates.  We  present the log  of all these  observations in
Table~\ref{table_swift_obs}.   Hereafter,  we  refer  to  each  single
observation, using only  the last two digits of  its ObsId.  We focus,
in the  present work, on the  data analysis and  interpretation of the
X-ray spectrum using data  from the Swift/XRT telescope, filtering out
time intervals around the  type-I X-ray bursts \citep[present in ObsId
32, 33 and 34; see][]{linares10}.

Data  were extracted  according  to standard  selection and  reduction
criteria \citep{burrows05} and  the latest available calibration files
(CalDB v.  20100528).  The Swift/XRT data were analysed in both WT and
PC  modes   (processed  with  the   \textit{xrtpipeline}  v.  0.12.3).
Filtering   and  screening   criteria  were   applied  by   using  the
\textit{ftools}  software package  (HEASoft v.   6.11).   We extracted
source  and background  light curves  and spectra  by  selecting event
grades  of 0--2  and  0--12 for  the  WT and  PC modes,  respectively.
Exposure maps  were created through the  \textit{xrtexpomap} task, and
we   used  the   latest  spectral   redistribution  matrices   in  the
\textit{HEASARC}  calibration  database.   Ancillary  response  files,
accounting  for  different  extraction  regions, vignetting,  and  PSF
corrections  were generated  using the  \textit{xrtmkarf}  task.  When
required, we corrected PC  observations for pile-up following standard
procedures.  The Swift/XRT light  curves were corrected for PSF losses
and vignetting using the \textit{xrtlccorr} task.

To perform the  SWIFT/XRT spectral analysis, we used  the 1.5--9.0 keV
energy range  for spectra in the  WT mode, as this  range provides the
highest  quality statistics  and calibration  accuracy.  We  note that
data  below 1.5  keV in  this  mode are  affected by  a spurious  soft
excess, that  has been previously  reported in case of  bright, highly
absorbed sources \footnote{$N_{\rm H}$ $\gtrsim$\,10$^{22}$ cm$^{-2}$,
see  \url{http://heasarc.gsfc.nasa.gov/docs/heasarc/caldb/swift/docs/xrt/SWIFT-XRT-CALDB-09_v16.pdf})}.
For  spectra taken in  PC mode,  we exploited  a broader  energy range
(0.8--9.0 keV),  as no  calibration issue affects  the softer  band in
this mode.   We rebinned the  spectra to have  at least 50  counts per
energy  channel  for spectra  with  high  quality  statistics; for  PC
spectra with  a small number  of total counts, channels  were rebinned
less coarsely with  a sampling of 25 counts per  channel, to apply the
$\chi^2$ statistics, still preserving  a sufficient number of channels
to allow a reliable estimate of the spectral shape.

\begin{figure}[ht!]
   \centering
   \includegraphics[angle=-90,width=\columnwidth]{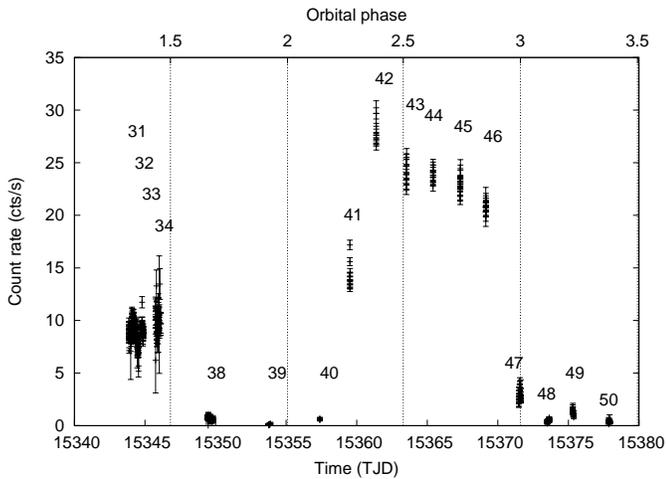}
   \caption{Light curve (time bin of 200 s) of the Swift/XRT observations of Cir X-1, with 
    indications of the   ObsIds  of   Table~\ref{table_swift_obs}.}
    \label{swift_lc}%
    \end{figure}

As shown in  Fig.\ref{asm_countrate}, the Swift observations partially
overlap with  the RXTE observations  during the declining part  of the
May outburst.   A close-up of  the Swift/XRT observations  with labels
indicating  the time  positions of  the different  ObsId, is  shown in
Fig.~\ref{swift_lc}. The Swift/XRT  data have complementary advantages
with respect  to the  PCA data,  as a higher  energy resolution  and a
softer  energy band.   However, its  effective area  decreases sharply
above 7  keV so  that the continuum  emission, determined by  means of
spectral model fitting, can  be difficult to constrain.  To compensate
for this,  we model the spectra,  when a quite close  (within one day)
RXTE observation is at hand, using both the Swift/XRT and the RXTE/PCA
data. The PCA  data are used in the 7.0--20.0 keV  range, to provide a
reliable  high-energy determination  of  the spectrum.   We take  into
consideration a relative  normalisation constant between the Swift/XRT
and RXTE/PCA  spectra to take  into account possible  flux differences
and calibration issues.

ObsId 31, 32,  and 33 were performed before  the slope change observed
in the light  curve of the RXTE  data; ObsId 34, 38, and  40 track the
rapidly decaying stage of the  May outburst to a very faint luminosity
state.  Starting from ObsId 41,  a new outburst was detected.  The new
outbursting phase lasted  for about half orbital cycle  and turned off
starting from the phase-zero passage  at TJD 15371.6.  The final three
observations cover the return to the pre-outburst, faint state.

%%%%%%%%%%%%%%%%%%%%%%
%%%%%%%%%%%%%%%%%%%%%

\subsection{The Chandra observation} \label{chandra_sect}

The  Chandra observation  started  on  2010 July  4  05:05:10 TT  (TJD
15381.212) and ended on 2010 July 4 11:10:20 TT (TJD 15381.466), which
correspond to  the phase interval 3.5823--3.5977. The  net exposure of
the observation  amounts to 20\,020  s. Chandra observed Cir  X-1 with
the  High Energy  Transmission Grating  Spectrometer (HETGS)  in FAINT
mode; we used Sim-Z=-8 mm  and a sub-array (ROWS=1--350).  The ACIS-S0
and ACIS-S5 were turned off in order to mitigate the possible presence
of  pile-up, given  a large  uncertainty in  the flux  prediction. The
count rate  in the HETGS was  however lower than  foreseen and pile-up
did not affect the first-order grating spectra.  The HETGS consists of
two types  of transmission gratings,  the Medium Energy  Grating (MEG)
and the High Energy Grating (HEG).  The HETGS provides high-resolution
spectroscopy  from 1.2  $\AA$  to  31 $\AA$  (0.4--10  keV); the  peak
spectral resolution is  of $\lambda / \Delta \lambda  \sim 1000$ at 12
$\AA$ for the  HEG first order.  The frame  time for this observation,
corresponding to the time resolution at which events are collected, is
1.24104 s.

\begin{figure}[ht!]
  \centering
    \includegraphics[width=\columnwidth]{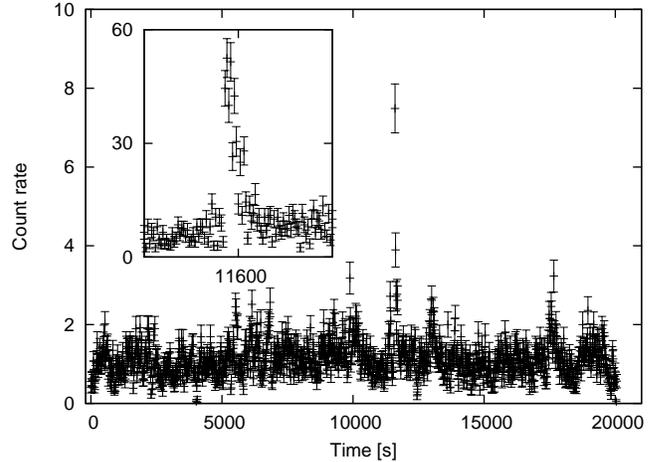}
    \caption{Chandra  light-curve  (sum  of first  order  HEG and  MEG
      counts) of the Cir X-1 observation. Time zero corresponds to TJD
      15381.21190.  Bin  time is 20  s. Insert panel: snapshot  of the
      type-I X-ray burst, bin time is 2 s. }
    \label{chandra_obs}%
\end{figure}

The   light   curve  of   the   Chandra   observation   is  shown   in
Fig.~\ref{chandra_obs}. During  the observation, a  type-I X-ray burst
is clearly detected 12\,930 s  after the start of the observation (see
Fig.~\ref{chandra_obs},  left  panel).   The  burst  shows  a  typical
fast-rise exponential-decay  (FRED) shape, with  a rise time $<$  2 s,
and an exponential decay time of 10.2 s.  The total energy released by
the burst is $\sim$ 2 $\times$ 10$^{38}$ erg.

In our spectral analysis, we  used the Chandra HEG and MEG first-order
spectra.  We examined only  the persistent  emission and  excluded the
time interval around  the burst event (10 s before the  peak and 100 s
after the peak).  We processed the event list using available software
(CIAO  ver.  4.1.2)  and computed  aspect-corrected exposure  maps for
each  spectrum in  order to  take into  account their  effects  on the
effective  area of  the  CCD spectrometer.  Spectra  were rebinned  to
ensure that we had at least 25 counts per channel energy.  We used HEG
(and  MEG)  data in  the  1.5--9.0 keV  (1.4--5.0  keV)  range, ans  a
normalisation multiplicative constant between  the MEG and HEG data to
take  into account  the relative  flux and  calibration uncertainties,
whose value was  0.94\,$\pm$\,0.4 for the HEG with  respect to the MEG
data-set (frozen to unity) for the model in Table\ref{specresults}.

%%%%%%%%%%%%%%%%%%%%%%%%%%%%%%%%%%%
%%%%%%%%%%%%%%%%%%%%%%%%%%%%%%%%%%%
%%%%%%%%%%%%%%%%%%%%%%%%%%%%%%%%%%%

\section{Spectral analysis} \label{sec_analisi}
\subsection{Model}

In   our   spectral   analysis,   we   used   Xspec   version   12.5.0
\citep{arnaud96}.   The fluxes and  associated errors  were calculated
using the convolution model  \texttt{cflux} available in Xspec.  Since
\texttt{cflux} calculates the flux only within the band covered by the
response matrix of the instrument,  we used a dummy response matrix to
calculate the  extrapolated luminosities.  Errors are  reported at the
90\% confidence level for a single parameter of interest.

We modelled the  spectra, in all the outburst  phases, using a thermal
Comptonized  component  \citep[\texttt{comptt}, ][]{titarchuk94}.  The
effect of interstellar absorption  ($N_{\rm H}$) was modelled with the
\texttt{phabs}  component.   Although  different two-component  models
(e.g. two black-bodies, or a black-body and a cut-off power-law) could
also provide  acceptable $\chi^2$ values, this  choice offered several
more solid advantages: it is  a model that closely matches the results
of previous  spectral decomposition attempts for  the available fitted
X-ray range; it avoids non-physically high extrapolated fluxes outside
the  range covered  by  the data;  it  allows us  to  follow the  most
significant spectral changes keeping  the number of free parameters to
a  minimum;  and  it  provides  statistically  acceptable  fits.   The
Comptonization is  determined by the  temperature of the  soft photons
($kT_0$),  the  electronic  temperature  ($kT_e$), the  optical  depth
($\tau$)  of  the cloud,  which  is assumed  to  be  spherical, and  a
normalisation  parameter related  to  the flux.   No other  additional
broadband continuum component is statistically required for any of the
examined spectra.

\begin{figure}[ht!]
   \centering
\includegraphics[angle=-90,width=\columnwidth]{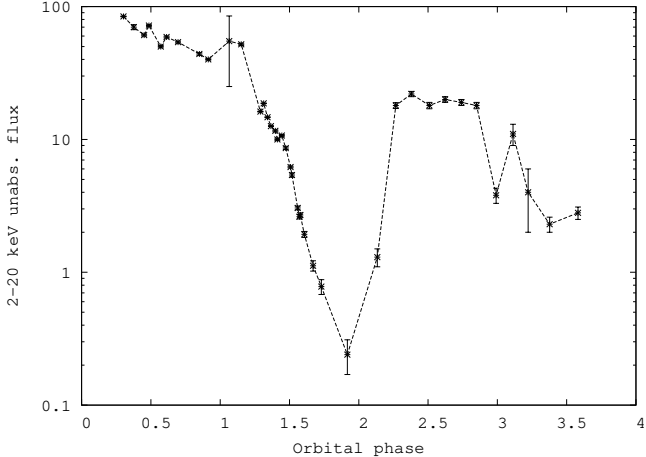} 
\caption{2--20 keV unabsorbed flux as a function of the orbital phase.}
    \label{flux_unabs}%
\end{figure}

In many  of the RXTE  spectra, the interstellar absorption  was poorly
constrained, when  \nh $<$ 2 $\times$\,10$^{22}$  cm$^{-2}$.  In these
cases, we set this parameter to 1\,$\times$\,10$^{22}$ cm$^{-2}$ (this
choice   will   be  motivated   a   posteriori,   see  discussion   in
Sect.~\ref{discussion1}),   to   better   constrain  the   other   fit
parameters.    We   added  a   neutral   partial  covering   component
(\texttt{pcfabs} in  Xspec), if the probability  of chance improvement
was less than  5\%.  When the \texttt{pcfabs} component  was used, the
\texttt{phabs} $N_{\rm H}$ value was  frozen to the reference value of
1\,$\times$\,10$^{22}$  cm$^{-2}$ to  avoid strong  correlations among
the  spectral  parameters.  This  component  was  required mostly  for
spectra  close to  the periastron.  The RXTE  observation  ObsId 03-01
displays this strong variability,  and we dedicate the next subsection
to this particular data-set.   In addition, an iron K$\alpha$ Gaussian
line, in  the 6.4--7.0 keV range,  was added to the  best-fit model if
its   significance   was  more   than   4   $\sigma$.    We  show   in
Table~\ref{specresults} the spectral  fitting results. Parameters with
no  associated  error  were  kept  frozen  during  the  fitting.   The
unabsorbed fluxes were calculated in  the 2.0--20 keV band in units of
10$^{-10}$ erg  cm$^{-2}$ s$^{-1}$, and  are plotted as a  function of
the orbital phase in Fig.\ref{flux_unabs}.

\subsection{Results}

Almost the entire May outburst is well-covered by RXTE pointings, with
a total exposure  of $\sim$ 74 ks in 23.7 days.   We collected a total
of 27 PCU2 spectra.  In 13 spectra, a Gaussian Fe K$\alpha$ line, with
rest-frame energies  of between  6.4 keV and  6.6 keV was  required to
obtain acceptable  $\chi^2$ values. The equivalent width  of this line
varied between 20 and 80 eV,  during the initial phase of the outburst
and showed a significant increase  as the spectrum became harder.  The
first  spectrum (ObsId  01-00)  is  the most  luminous  of the  series
(8.4\,$\times$\,10$^{-9}$ erg cm$^{-2}$ s$^{-1}$), showing evidence of
local  neutral absorption  at orbital  phase 0.302.   The  spectrum is
satisfactorily described  by an optically  thick Comptonized component
($\tau$  $\sim$\,12),  where  soft  photons of  $\sim$\,0.84  keV  are
upscattered by thermal electron  of energy $\sim$\,2.74 keV.  No other
component  is statistically required,  given the  generally acceptable
$\chi^2$ values. The following hardening  of the spectrum (region B of
Fig.\ref{rxte_hid}) reflects the gradual increase in optical depth and
the  almost constant  value  of the  electron  temperature.  The  soft
temperature  varies  between  0.74   keV  and  1.07  keV.   After  the
periastron passage  (ObsId 03-02), the  spectrum shows an  increase in
luminosity and a  high column density (region C).  As the source moves
towards the  apastron (region D),  the luminosity drops,  returning to
the pre-periastron trend of  linear decrease, and the local absorption
decreases, thus giving the softening  observed in region D, whereas in
the high-energy band no significant shape variation is observed.

The final stage  of the outburst observed with  RXTE (region E), after
the  ankle   in  the  light  curve,  encompasses   the  orbital  phase
1.56--1.73.   To increase the  statistics, all  the PCU2  spectra were
summed together  and fitted together  with the Swift-XRT  spectrum 38.
We verified that the PCU2 spectra were all compatible with each other,
excluding  only  the  expected   variation  in  the  fluxes  that  are
separately reported in  Table~\ref{specresults}.  No significant local
absorption is detected in the  spectrum, but there is an indication of
a broader fluorescence line Fe  K$\alpha$ line with respect to that in
the soft state.  The most interesting aspect is the sudden jump in the
electron temperature  to values above 20 keV,  occurring between phase
1.51 and  1.55, which  is consistent with  the apastron  passage.  The
spectral change  in the  electron temperature/optical depth  turns the
Comptonizing  corona from  an  optically thick  to  an optically  thin
regime.   This spectral  variation is  statistically  very significant
(fixing the electron temperature at  2.7 keV would result in a reduced
$\chi^2$ of  6.3), and the lower  limit, at 95\%  confidence level, is
found at 14 keV. Since the data  cover the spectrum only up to 20 keV,
we chose to  fix this parameter to this boundary  value.  We note that
the passage  to the optically thin  regime still requires  a source of
soft  photons at a  relatively high  temperature.  The  probability of
chance  improvement using  an F-test  for the  addition of  the $kT_0$
parameter  is $\sim$\,0.3\%.   The luminosity  at  which the  spectral
transition  (ObsId 04-08)  takes place,  considering  the extrapolated
0.1--100  keV  range   is  (3.5\,$\pm$\,0.7)  $\times$\,10$^{36}$  erg
s$^{-1}$.

Swift ObsId  39 and  40, taken during  the second  periastron passage,
mark the end of the  May outburst since the source subsequently passed
to   a  very   faint   state  (corresponding   to   a  luminosity   of
$\sim$\,6\,$\times$\,10$^{35}$  erg   s$^{-1}$).  Owing  to   the  low
statistics, we were unable  to constrain the broadband spectral shape;
if we impose the same model of ObsId 38, we determine a value of local
absorption  $N_{\rm   H}$  $\gtrsim$\,10$^{23}$  cm$^{-2}$   for  both
observations.

Another outburst is  then observed between phases 2.14  and 2.27.  The
rising phase is characterised again  by a soft spectrum and moderately
high local  $N_{\rm H}$ in the  (4--5.5) $\times$\,10$^{22}$ cm$^{-2}$
range (ObsId 41).   The rising time of this  outburst is comparable to
that of  the May outburst (about  two days).  When  the source reaches
the  maximum peak  intensity (ObsId  42), the  spectral  shape remains
substantially unaltered up to the next periastron passage. Because the
data  cover only the  soft X-ray  range, we  were unable  to constrain
simultaneosly  the electron  temperature and  optical depth.  For this
reason, we  chose to fix the former  to a reference value  of 2.6 keV.
Quite interestingly, the spectrum and the flux appear to be consistent
with  the spectrum  at the  previous same  orbital phase  (e.g.  ObsId
03-03).  However,  in contrast  to that case,  there is  no subsequent
clear decrease in  the flux; the flux remains at  a value of $\sim$\,2
$\times$\,10$^{-9}$ erg cm$^{-2}$ s$^{-1}$ for the following ten days.

\begin{figure*}[ht!]
   \centering
\begin{tabular}{cc}
 \includegraphics[angle=-90,width=0.85\columnwidth]{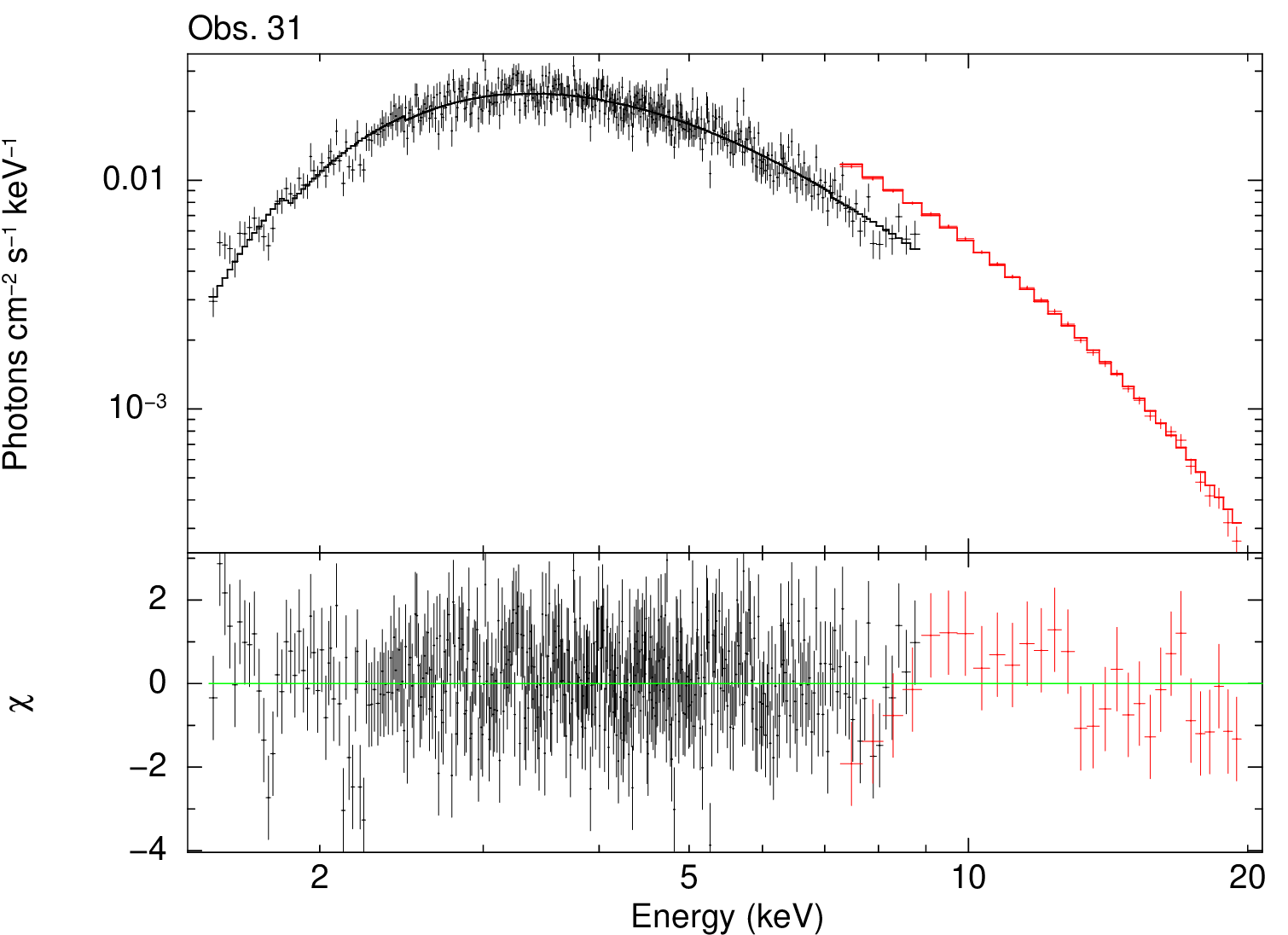} & 
\includegraphics[angle=-90,width=0.85\columnwidth]{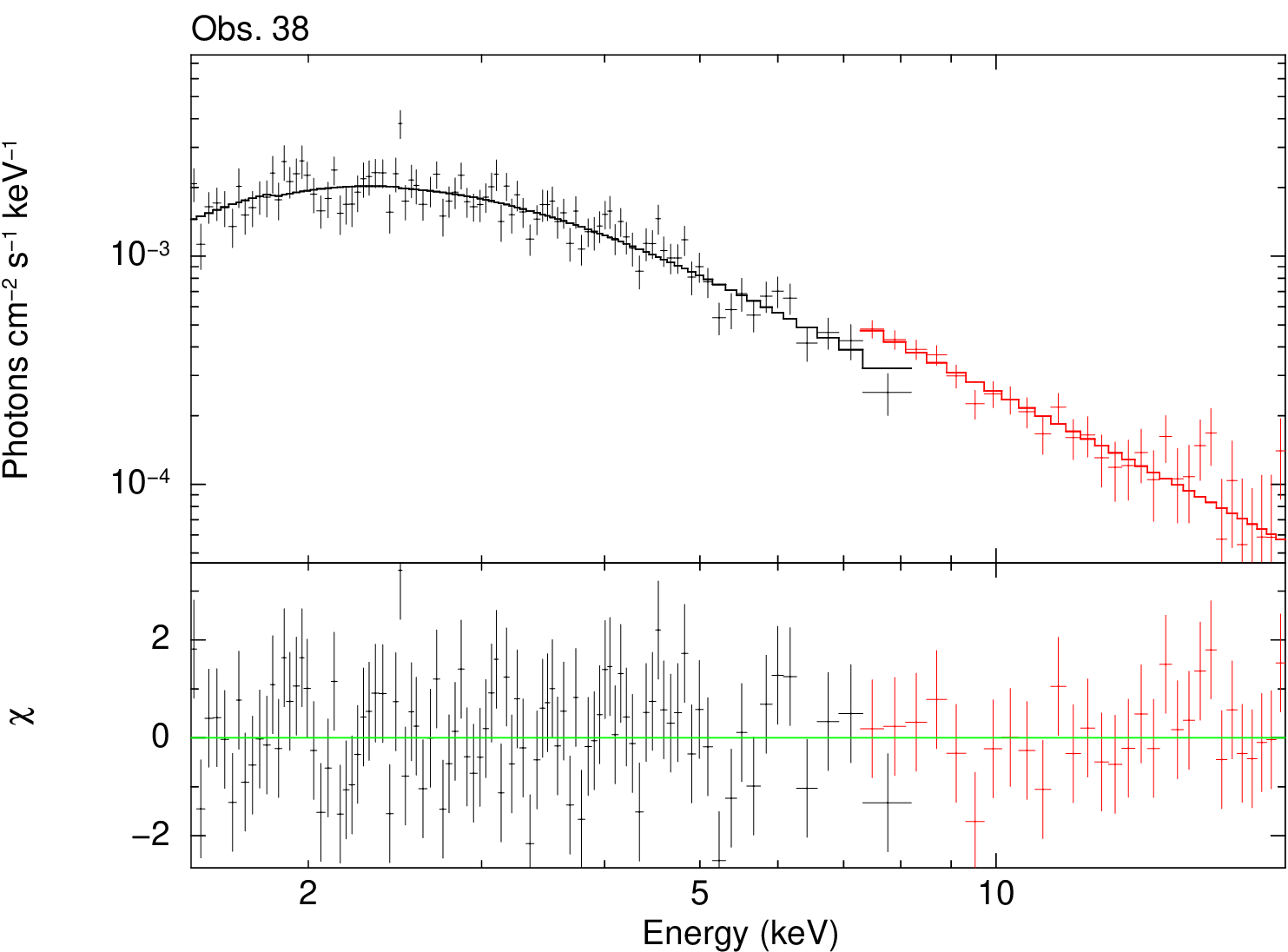}\\
 \includegraphics[angle=-90,width=0.85\columnwidth]{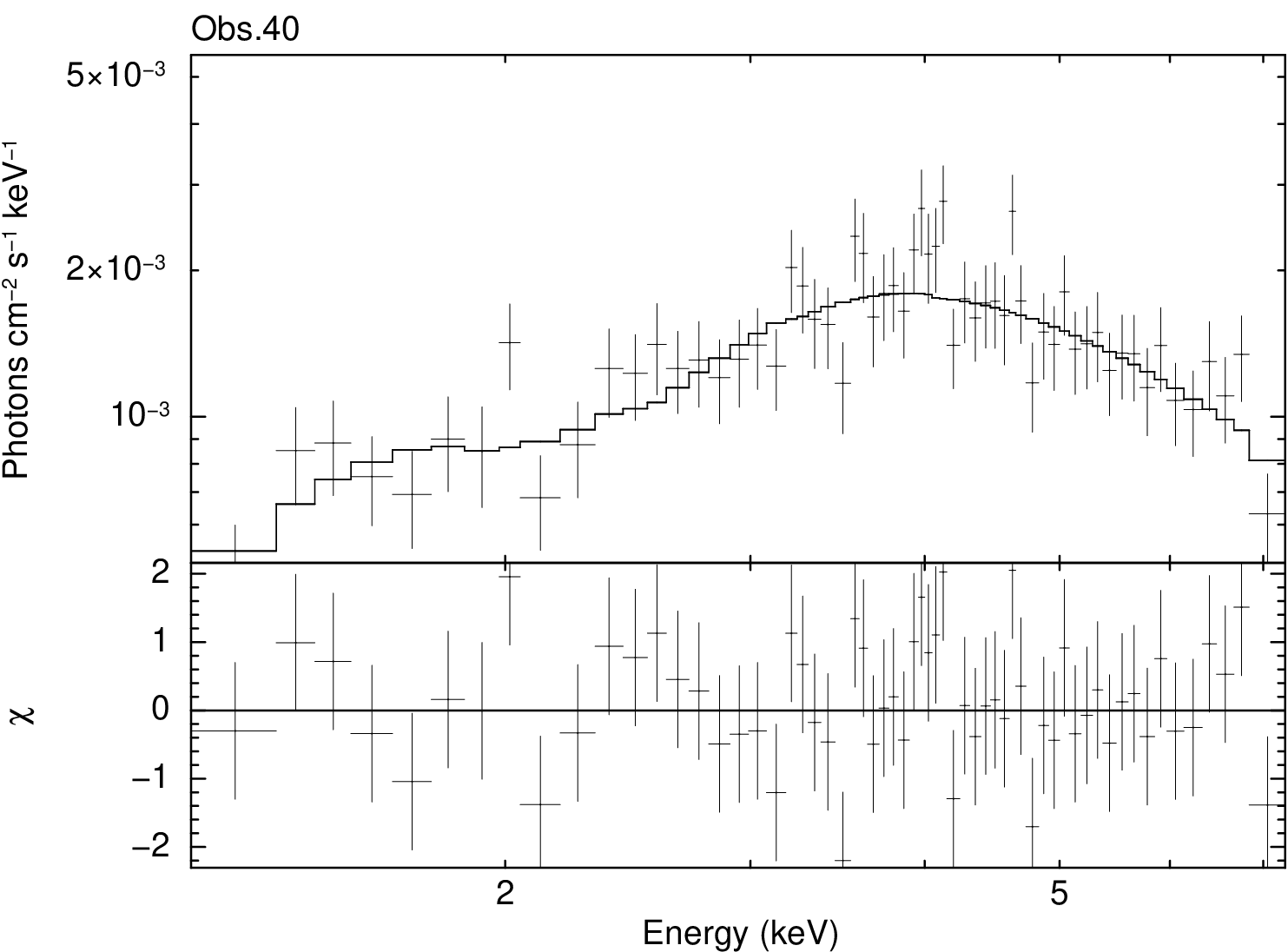} & 
\includegraphics[angle=-90,width=0.85\columnwidth]{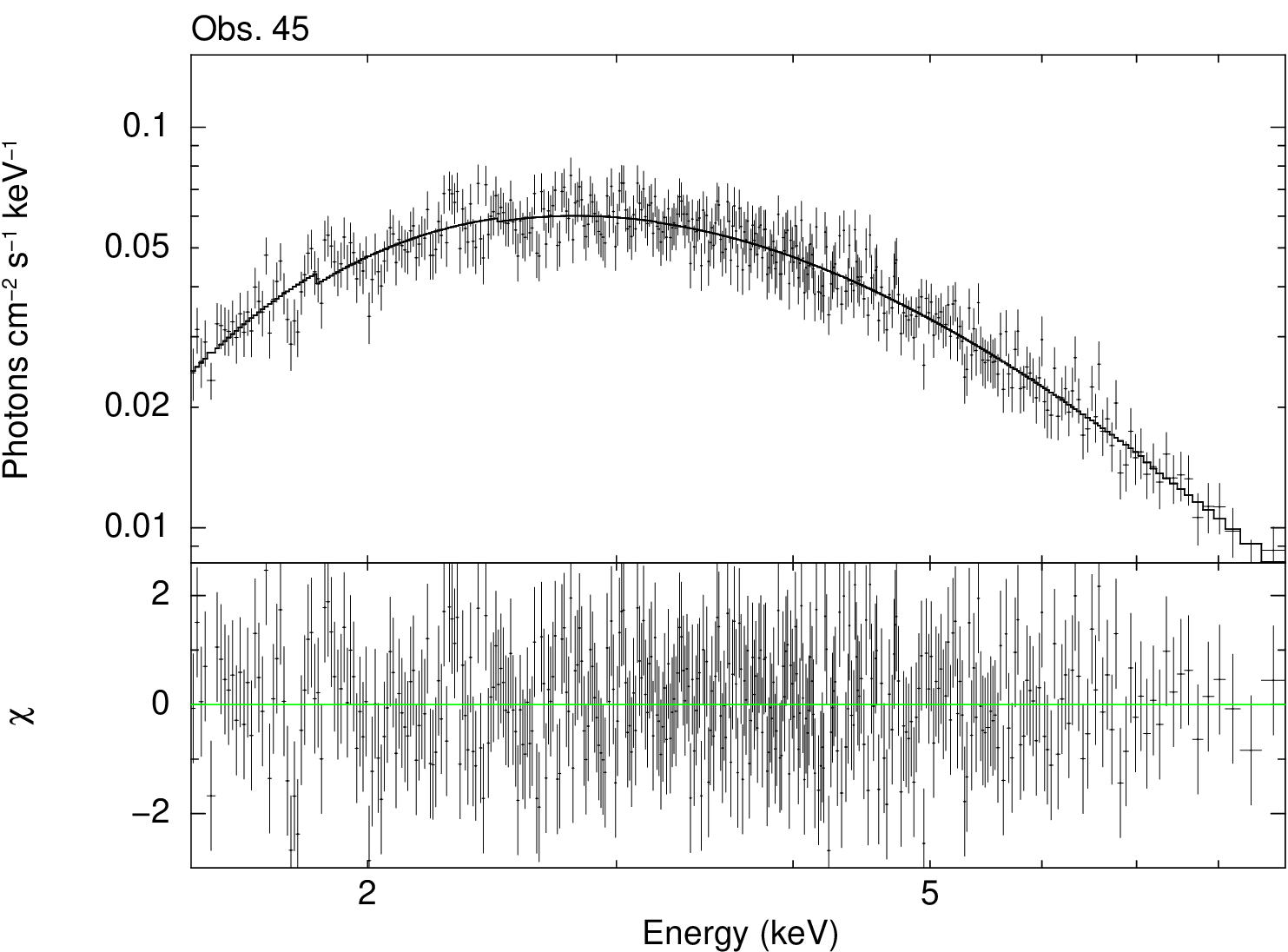}\\
 \includegraphics[angle=-90,width=0.85\columnwidth]{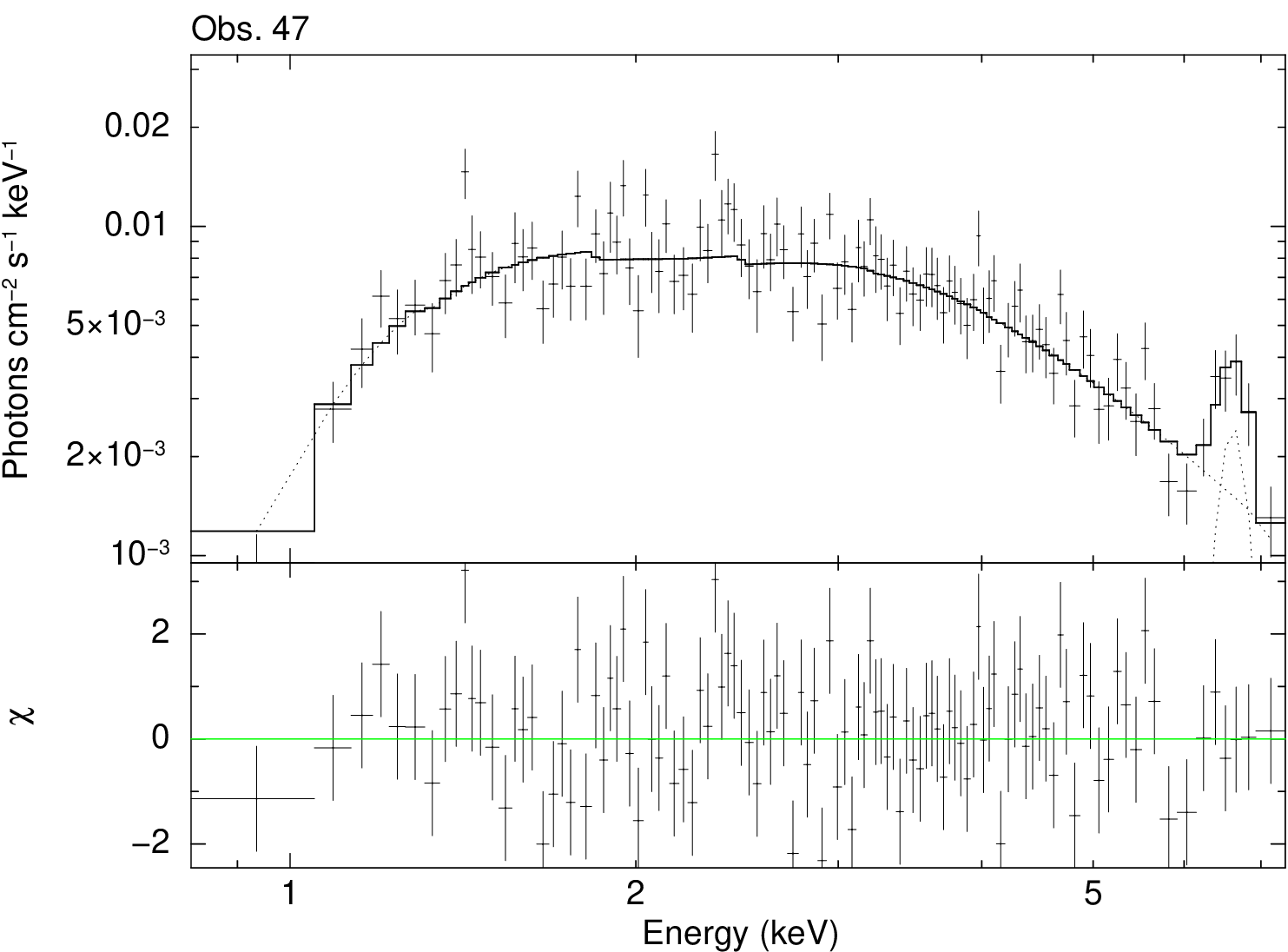} & 
\includegraphics[angle=-90,width=0.85\columnwidth]{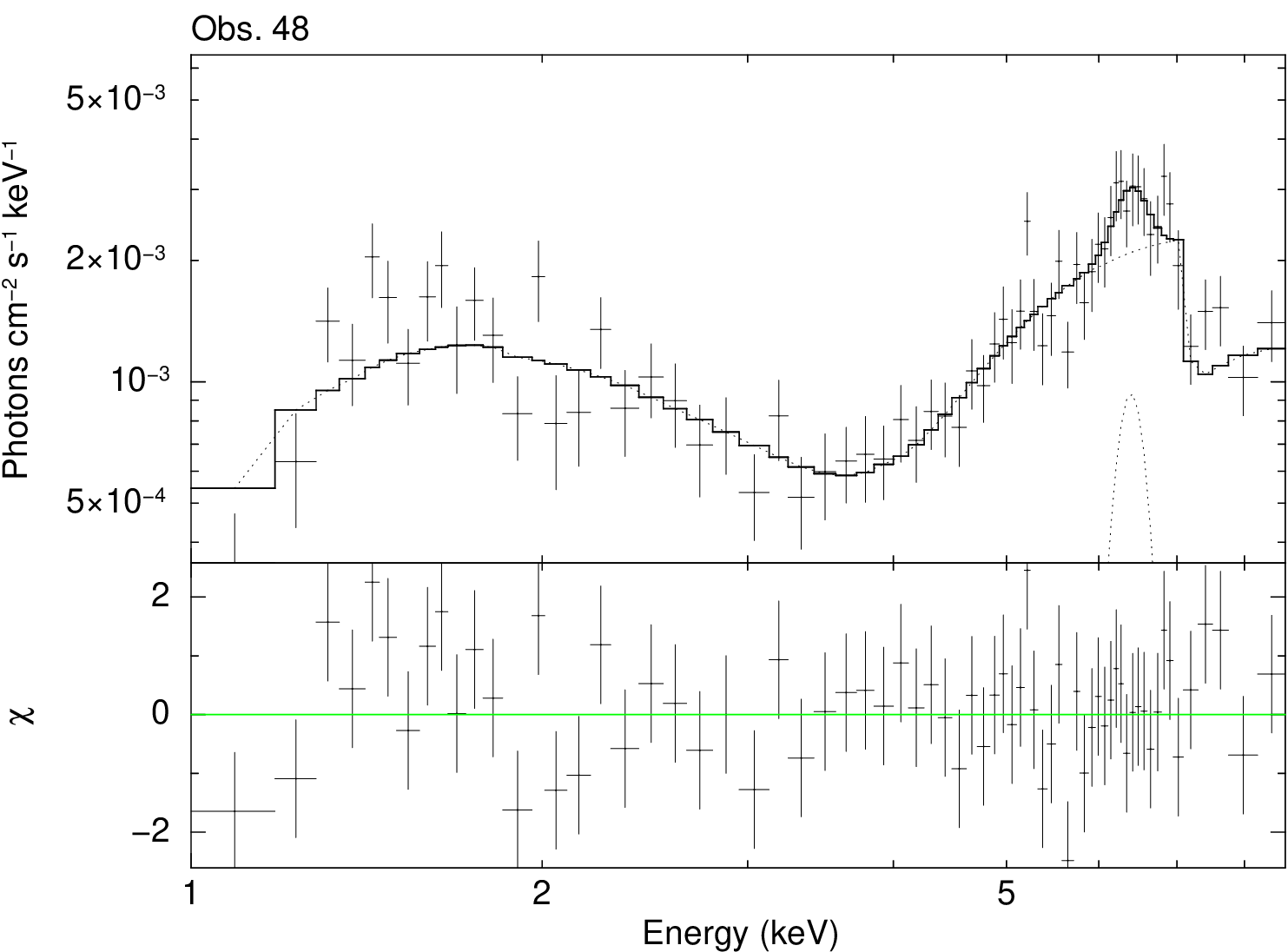}\\	
\end{tabular} 
  \caption{Plot   of  unfolded   spectra,  best-fit   models,  and
    residuals in  units of  $\sigma$ for six  representative
    spectra. ObsId 31 and ObsId 38 also show, in red, RXTE/PCA data used
    to  more tightly constrain  the fitting  model.}
    \label{swift_spectra}
\end{figure*}

During ObsId  47, the source  is precisely at the  periastron passage.
The spectrum again  shows an increase in local  absorption and partial
covering of the  direct emission. This passage seems  to suddenly turn
off the ongoing outburst, as can be more clearly seen by the long-term
ASM light  curve (Fig.~\ref{asm_countrate}).  We  cannot discriminate,
based  on the  goodness of  the fit,  between a  Comptonized  model of
either  high  or  low  electron  temperature, so  that  we  report  in
Table~\ref{specresults}, both spectral solutions with fixed values for
the electron  temperature at  2.6 keV (corresponding  to a  soft state
model) and  20 keV  (corresponding to a  hard state model).   ObsId 48
shows the source in a dipping state, following the periastron passage.
Local  absorption reaches  6\,$\times$\,10$^{23}$  cm$^{-2}$, and  the
source flux  shows a  significant rise. A  very intense iron  line, of
$\sim$\,0.7 keV  equivalent width, is  also detected in  the spectrum.
After this dip phase, the ObsId 49 and 50 indicate that there is still
significant  local absorption (\nh  $\sim$\,(2--3) $\times$\,10$^{23}$
cm$^{-2}$),   but  the  source   flux  decreases   considerably.   The
Fig.~\ref{swift_spectra} shows the unfolded models and spectra for six
different  phases,  during  and  amid  the  two  outbursts,  that  are
representative  of   the  spectral   evolution  of  the   source  (see
Table\ref{specresults}).

%%%%%%%%%%%%%%%%%%%%%%%
%%%%%%%%%%%%%%%%%%%%%%%

\begin{figure}[ht!]
   \centering
\includegraphics[angle=-90,width=\columnwidth]{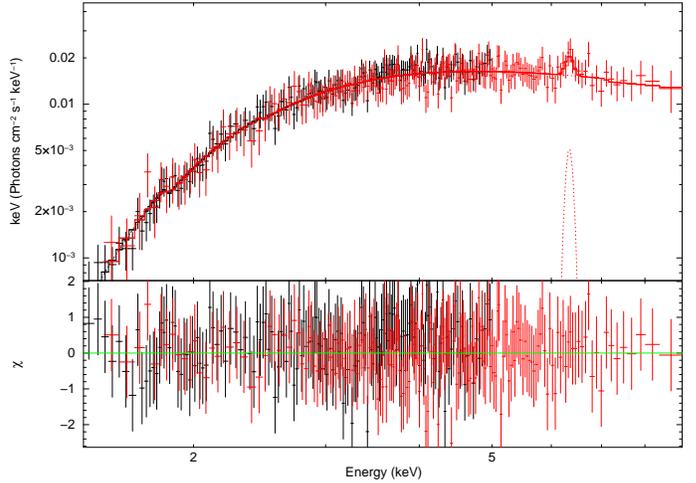} 
\caption{Unfolded data  with   best-fit   model and residuals in units
of $\sigma$ for the Chandra observation. HEG data in red, MEG data in black.}
    \label{chandra_models}%
\end{figure}

The  spectral analysis  of the  20 ks  Chandra observation  reveals an
emission feature, whose  energy is constrained to be  between 6.30 keV
and 6.44 keV, which we identify as a Fe K$\alpha$ fluorescence line of
neutral,  or  mildly ionised,  iron.   The  line  is clearly  detected
(F-test for chance improvement $\sim$ 1\%), with an upper limit to its
width ($\sigma$) of 0.2 keV.   The shape of the continuum emission can
be  described to  first order  by an  absorbed (\nh  = 4.3\,$\pm$\,0.3
$\times$\,10$^{22}$     cm$^{-2}$)     power-law     (photon     index
$\Gamma$\,=\,1.46\,$\pm$\,0.11),   giving   already   a   satisfactory
description of  the data ($\chi^2$\,=\,372, 665  d.o.f.). The presence
of a high-energy  cut-off is not statistically required,  as a cut-off
power-law only marginally improves the fit ($\Delta \chi^2$\,=\,3, for
one  d.o.f.   more).  However,  to  be  consistent  with the  previous
spectral  model,  we also  show  in Table~\ref{specresults},  spectral
fitting results  adopting the $kT_e$\,=\,20 keV  model constraint.  In
Fig.\ref{chandra_models},  we  show  the  unfolded  model,  data,  and
residuals.

The  0.5--10   keV  unabsorbed  flux   (2.4\,$\times$\,10$^{-10}$  erg
cm$^{-2}$ s$^{-1}$) is close to the value obtained for the Swift ObsId
50, but the  amount of local absorption is a factor  of ten less.  The
extrapolated  X-ray flux  in  this state  is  difficult to  constrain,
because   of  the   uncertainty   in  the   cut-off   energy  of   the
Comptonization.  A value  $\sim$\,5 $\times$\,10$^{-10}$ erg cm$^{-2}$
s$^{-1}$ appears however, quite reasonable, because a ratio of two for
the 0.1--100 keV  range to the 0.5--10 keV range,  is derived from the
fitting of  the spectra  in the region  E of  Fig.\ref{rxte_hid}.  The
corresponding  isotropic  luminosity  is 3.5\,$\times$\,10$^{36}$  erg
s$^{-1}$.

In this faint state, a  typical type-I X-ray burst is clearly observed
in the light curve.  The X-ray bursts observed during the May outburst
were always associated with the soft state, when the source luminosity
varied   between  1.4   and   5.9\,$\times$\,10$^{37}$  erg   s$^{-1}$
\citep{linares10}.   To  make  a  rough  guess of  the  ratio  of  the
time-averaged persistent  to burst luminosity  ($\alpha$), we consider
here only  the collected counts during  the burst, as  the low quality
statistics do not  allow us to perform a  sensible time-resolved burst
spectral analysis.   If the time  between bursts is  approximately the
time duration of the  Chandra observation ($\Delta t \approx$\,20 ks),
this leads  to $\alpha \approx$\,180.   This value is  consistent with
the values derived for  some of the previous bursts \citep{linares10},
and points  to a pure  He burst, as  expected from the  low persistent
accretion rate \citep{narayan03}.

\subsection{Passage during dips: RXTE observation 03-01}
\label{sec_obs0301}
\begin{figure}[ht!] 
  \begin{tabular}{c}
   \includegraphics[angle=-90,width=8cm]{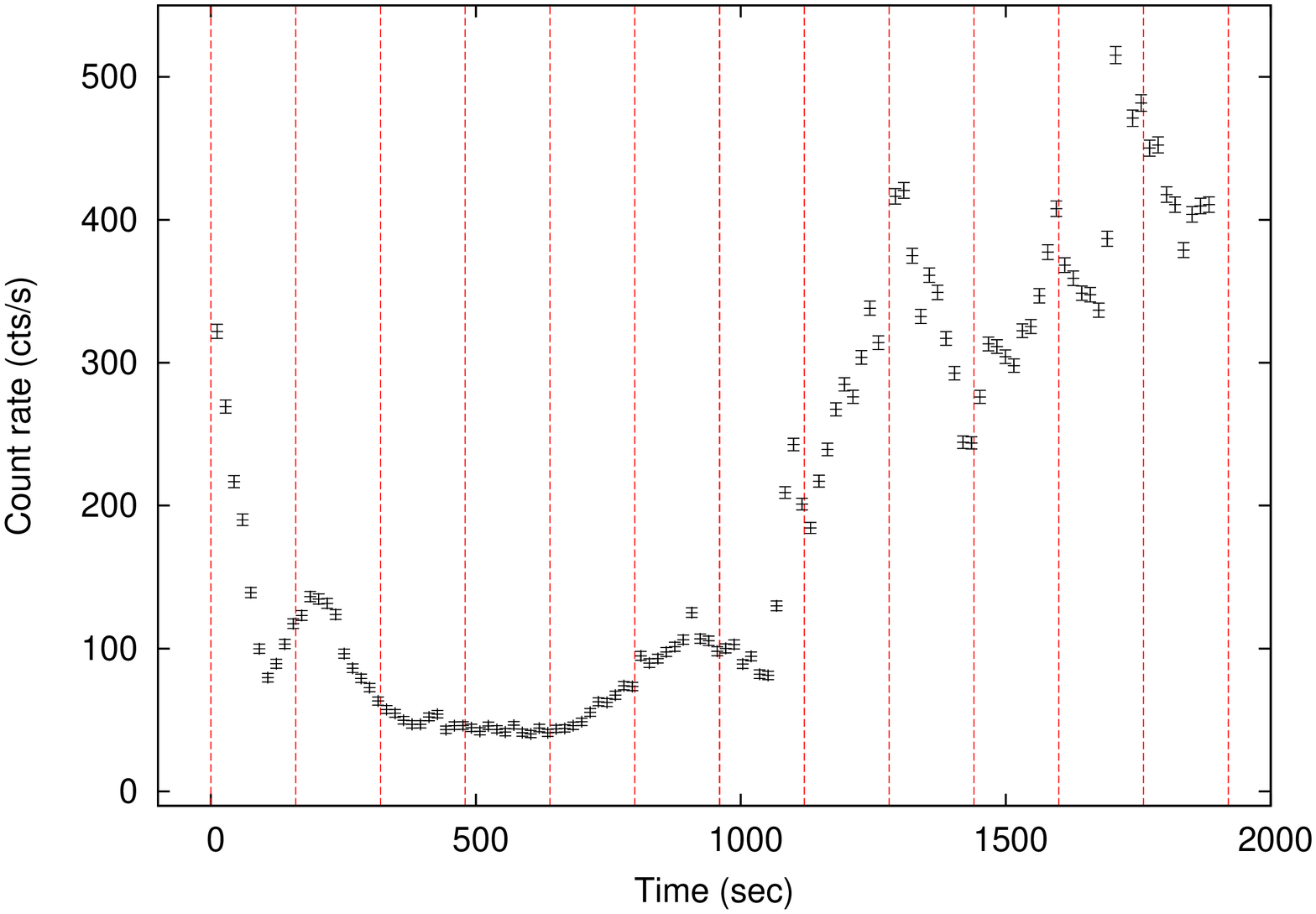} \\
   \includegraphics[angle=-90,width=8cm]{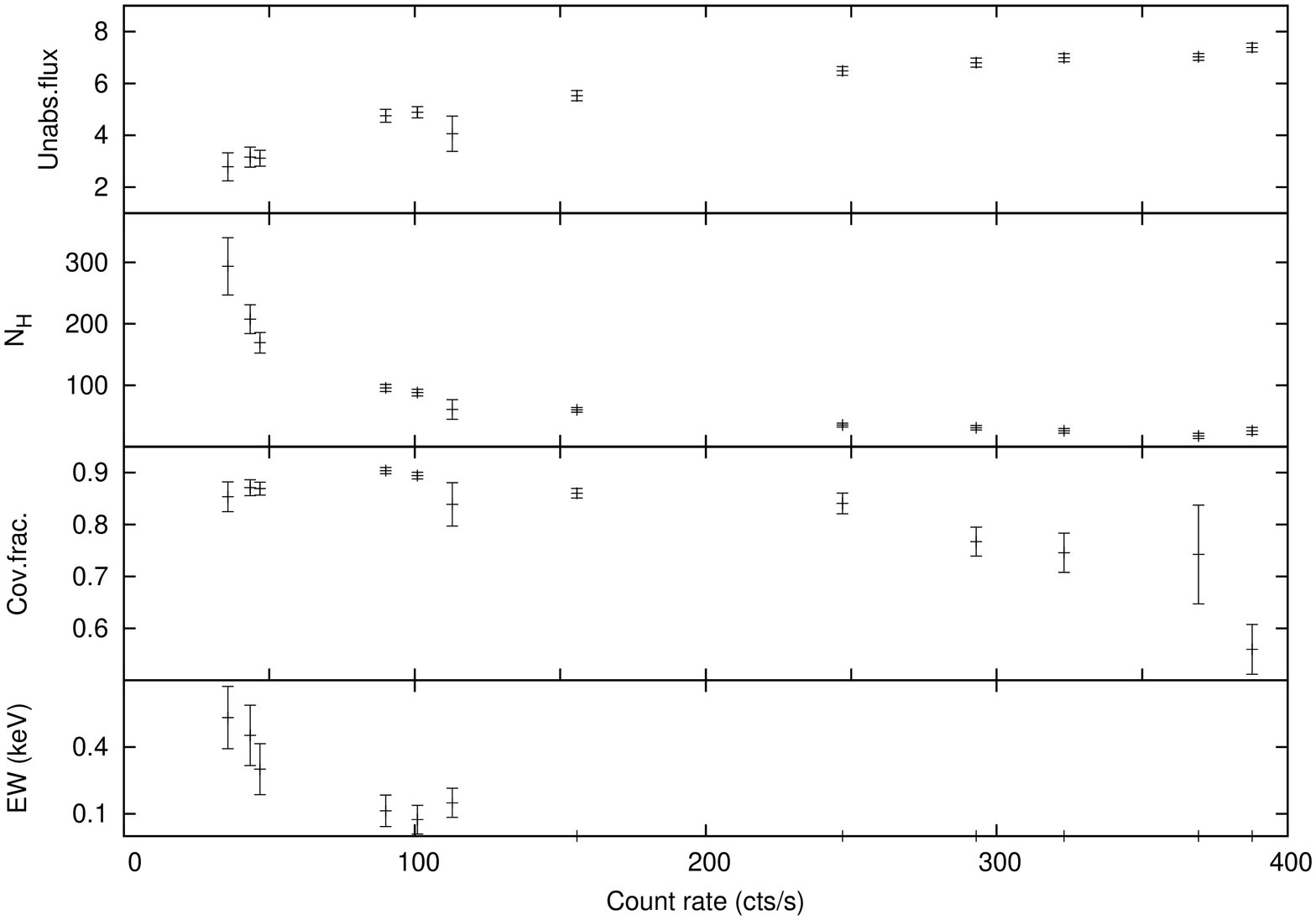} \\
    \includegraphics[angle=-90,width=8.5cm]{spectra0301_2.ps}
\end{tabular}   
   \caption{\textit{Upper panel:} net light curve of ObsId  03-01, showing the
     time   intervals  used  for   spectral  analysis.  \textit{Middle panel:}
     best-fit spectral results as a function of the observed count
     rate;  from top  to bottom:  unabsorbed flux  of  the Comptonized
     component   in  units  of   10$^{-9}$  erg   cm$^{-2}$  s$^{-1}$,
     equivalent  hydrogen column of  the \texttt{pcfabs}  component in
     units of  10$^{22}$ cm$^{-2}$, covering  fraction, and equivalent
     width  of the  Fe  K$\alpha$  Gaussian line (if  present in  the
     best-fit model). \textit{Bottom panel:} 
     Four unfolded  spectra from  ObsId 03-01.  From  top to
     bottom, unfolded spectra of time segments 11, 8, 2, and 5.}
 \label{figobs0301}
    \end{figure}

The  light curve  of RXTE  ObsId 03-01  in  Fig.\ref{figobs0301} shows
strong variability:  a dip  phase at the  start of the  observation is
then followed by multiple flare  episodes.  To track the fast spectral
variability for this observation, we divided the light curve into ten,
equally  spaced intervals.   This  choice is  a reasonable  compromise
between  the fast  tracking of  the X-ray  variability and  spectra of
sufficiently high  signal-to-noise ratio.  We used  a partial covering
component  (\texttt{pcfabs} in  Xspec)  to model  the  effects of  the
varying  absorption.  We  also take  into account  electron scattering
(\texttt{cabs} in Xspec),by  insisting that its value be  equal to the
\nh of the \texttt{pcfabs} component, as done in \citet{shirey99}.  We
kept the value  of the interstellar \nh fixed  at 1 $\times$ 10$^{22}$
cm$^{-2}$.  We used the previously discussed \texttt{comptt} component
to describe the continuum  emission.  Although changes to the spectral
shape  could  possibly  take  place,  we assumed,  as  a  first  order
approximation, that the  spectral shape did not change  during the dip
passage (by keeping the soft photon temperature, electron temperature,
and  optical  depth fixed  to  the  average  values found  during  the
pre-deep   observation,   i.e.   to   0.81   keV,   2.55  keV,   17.3,
respectively).  We  allowed the hydrogen column  density, the covering
fraction of  the \texttt{pcfabs}  component, and the  normalisation of
the \texttt{comptt}  component free  to vary in  each of  the spectral
fits.   In some of  the spectra,  especially during  the dip  phase, a
broad  Gaussian line  was  definitely required  to achieve  acceptably
small $\chi^2$ values.

This approach was successful in describing the spectral shape, with an
average reduced $\chi^2$ of 1.1 (for 34/35 d.o.f.).  In the left panel
of  Fig.\ref{figobs0301},   we  show  the  net  light   curve  of  the
observation  and in  the right  panel  the evolution  of the  spectral
parameters. The  dip phase is  characterised by \nh values  of between
1.5 and  3.5 $\times$ 10$^{24}$ cm$^{-2}$,  and corresponding covering
fractions of between 80\% and 90\%.  During this phase, the unabsorbed
flux  was   $\sim$  2  $\times$  10$^{-9}$   erg  cm$^{-2}$  s$^{-1}$,
corresponding to  an isotropic luminosity of a  1.4 $\times$ 10$^{37}$
erg s$^{-1}$.  In  this phase, a broad fluorescent  Gaussian iron line
was detected  in the spectra,  with equivalent widths  decreasing from
600  eV to  100 eV,  in correlation  with the  decrease in  the column
density value.  As the source moved  out of the dip, variations in the
local  \nh  tended  to  flatten,  remaining  however  above  10$^{23}$
cm$^{-2}$.  The rapid rise in the count rate, observed outside the dip
phase, is mainly due to the rapid decrease in the local absorption and
an increase  of a  factor $\sim$  3 in the  intrinsic flux,  while the
covering fraction  follows a smooth trend towards  an asymptotic value
of  $\sim$ 50\%.  In  the bottom  panel of  Fig.\ref{figobs0301}, four
representative unfolded spectra show different unfolded spectra.

\section{Discussion}

\subsection{The spectral evolution of Cir X-1 during the outburst} \label{discussion1}

A multi-satellite  campaign of Cir X-1  from May to  July 2010 tracked
the spectral  evolution of  the source in  one of its  major outbursts
after  a long  period of  relative X-ray  quiescence during  the years
2007-2010.  RXTE  began its monitoring  campaign of Cir  X-1 $\sim$1.5
days after the outburst's peak at orbital phase 0.3.  During the first
part of the RXTE  monitoring (regions A--D of Fig.\ref{rxte_hid}), the
spectral  shape  of Cir  X-1  could  be  closely reproduced  by  thick
Comptonized  spectrum,  where  the   average  values  of  soft  photon
temperature  (0.9 keV),  electron temperature  (2.6 keV),  and optical
depth (12.9)  displayed relative variations of between  10\% and 20\%,
describing a  progressive hardening of  the spectrum before  the first
periastron  passage (with  an increase  in  the optical  depth) and  a
softening  after  this  passage.   In \citet{iaria05},  the  broadband
(0.1--200 keV) X-ray  spectrum of Cir X-1 at  orbital phase 0.62--0.84
and a luminosity  comparable to the luminosity reported  for regions A
and B of Fig.\ref{rxte_hid}, was modelled with a soft black-body ($kT$
$\sim$  0.5 keV),  an optically  thick Comptonized  component ($kT_0$,
$kT_e$, $\tau$ $\approx$ 1.1 keV,  2.65 keV, and 12, respectively) and
a series of  absorption edges at low energies  of highly ionised ions.
We found  that the  cut-off in the  Comptonization, determined  by the
electron temperature and optical depth, was remarkably consistent with
values in  that spectrum.  In our  analysis, we were  unable to assess
the  presence of  either  an  additional softer  component  or a  warm
absorber, but  the value  we derived for  the soft  photon temperature
appears  to  be the  average  the  two  soft temperatures  (black-body
temperature and $kT_0$ value) found in \citet{iaria05}.

Except  for  the  first  observation,  ObsId  01-00,  ($N_{\rm  H}$  =
3.2\,$\pm$\,1.1  $\times$ 10$^{22}$  cm$^{-2}$),  and the  observation
following   the   dip  behaviour,   ObsId   03-02,   ($N_{\rm  H}$   =
5.1\,$\pm$\,0.5  $\times$  10$^{22}$ cm$^{-2}$),  all  the other  RXTE
spectra  are   consistent  with  a  relatively  lower   value  of  the
interstellar absorption  ($\lesssim$\,2 $\times$ 10$^{22}$ cm$^{-2}$).
It is interesting to question how low this value could effectively be,
as  it  is the  only  way  of constraining  the  extent  of the  local
absorber.  The XRT  data provide a far more  comprehensive coverage at
low energies than PCA  spectra.  From Table~\ref{specresults}, we note
that  ObsId 34 and  ObsId 38  have the  lowest column  densities, that
point  to a  possible  lower  limit value  of  1.0 $\times$  10$^{22}$
cm$^{-2}$, which  was therefore adopted  in the fits, when  either the
parameter could not be constrained or we had evaluated the presence of
a  partial covering  component\footnote{This  parameter (expressed  in
  units of 10$^{22}$ cm$^{-2}$), however, depends on the adopted value
  for elements cross-sections and abundances.  Fitting the XRT ObsId34
  spectrum (which  has the  highest quality statistics),  we obtained,
  using   the   \texttt{bcmc}   cross-sections   \citep{balucinska92},
  1.60\,$\pm$\,0.24     (abundances    from     \citet{wilms00}    and
  \citet{asplund09})    and    1.68\,$\pm$\,0.26   (abundances    from
  \citet{lodders03});    using   the    \texttt{vern}   cross-sections
  \citep{verner96},  we  obtained  1.66\,$\pm$\,0.24 (abundances  from
  \citet{wilms00}   and   \citet{asplund09}),  and   1.74\,$\pm$\,0.26
  (abundances from \citet{lodders03}).   Calculations using the LAB HI
  survey  \citep{kalberla05}  and  the  DL  database  \citep{dickey90}
  result  in an  average value  of 1.49  and 1.84,  respectively.  The
  value derived  from the interstellar column density  towards Cir X-1
  become compatible with that expected from the radio HI maps.}.

The  RXTE  ObsId  03-01  (Sect.\ref{sec_obs0301}) indicated  that  the
periastron passage  significantly alters the spectral  shape, which is
mostly  affected  by  the  variability  of  the  cold  absorber.   The
simultaneous presence of a ionised warm absorber could not be assessed
given the  lack of spectral resolution  \citep{dai07a}.  Absorption is
however incomplete,  and $\sim$ 10\%  of the primary  emission remains
unabsorbed.  This  behaviour, which  was previously observed  when the
source was  in a bright  persistent, state, depends  on time-dependent
local conditions of the  medium, where interactions with the companion
star  causes turbulent  star losses  that may  partially  overfill the
outer disk  region's of  the accreting source  \citep{ding06}.  During
this passage the continuum  flux also significantly varies and reaches
values comparable with the peak  flux of the outburst. It is, perhaps,
not  coincidental that  the dipping  phase is  followed by  a multiple
flaring episode,  with the flux  of the source  rising by a  factor of
about three in about 1\,000 s \citep[see also][]{dai07a}, so that what
is  actually accreted during  the flaring  could be  part of  the same
matter  that  caused the  dipping  \citep[see  also][for a  convincing
application of  this scenario]{bozzo11}.  However, since  the slope of
the   decay  remains   the  same   in  both   regions  B   and   C  of
Fig.\ref{rxte_hid},  and the spectral  shape in  these two  regions is
also consistent (apart from the softening and hardening trends that we
found), we conclude that the  transient accretion phase at the dipping
is restricted  only to the periastron  passage and does  not alter the
overall outbursting process.

When Swift began observing Cir X-1 at phase 1.32, the source was still
in  the   optically  thick  regime.    Combined  XRT  and   PCU2  fits
significantly  improved  the   determination  of  the  \texttt{comptt}
spectral parameters,  providing tight constraints on the  value of the
local absorber, that monotonically  decreased from phase 1.32 to phase
1.68. During  phases 1.52--1.62,  with a 2--20  keV flux  of $\sim$5.4
$\times$  10$^{-10}$ \ergsec, the  spectrum changed  significantly, as
the  Comptonization cut-off  energy abruptly  moved out  of  band. The
corresponding  luminosity was  $\sim$2\%  of the  Eddington limit  (we
assume $L_{\rm Edd}$\,=\,1.8\,$\times$\,10$^{38}$ ergs s$^{-1}$ for a 1.4
M$_{\sun}$  NS). These state  transitions are  also often  observed in
some persistent  and transient accreting NS low-mass  binaries at this
energy  threshold   \citep{gladstone07}.   The  peculiar   system  XTE
J1701$-$462, which is the only  transient NS that started its outburst
showing classical Z-type behaviour  and gradually passing to the atoll
state  at lower  accretion rates,  showed its  transition to  the hard
state at the same  Eddington luminosity \citep{homan10}.  At very high
accretion  levels,  Cir X-1  showed  timing  characteristics that  are
similar to the Z-class sources \citep{shirey99, boutloukos06}, as well
as a state   closer    to   the    hard/atoll   behaviour
\citep{oosterbroek95}.  The present  analysis shows the first evidence
of a clear  passage from the soft to the  hard state, constraining the
luminosity threshold at which it happens.

The state transition  is also visible as a change in  the slope of the
count  rate (flux)  versus time  plot  (Fig.~\ref{rxte_lc_fit}). These
features  (which are  mostly referred  to as  knees, as  the  slope is
mostly  observed to  steepen) are  widely observed,  e.g.   during the
final  stages   of  the  outbursts  of  accreting   ms  X-ray  pulsars
\citep[defined  as \textit{brinks}  by][  when the  decay passes  from
exponential  to linear]{powell07},  but no  evident  and corresponding
change  in the  spectral  shape is  generally  observed. The  physical
mechanism behind these  changes in slope may therefore  not be unique;
while the onset of a  quasi-propeller phase is reasonable for the case
of accreting ms pulsars, whose  accretion flow can be channeled by the
NS magnetic field,  as discussed in \citet{hartman11}, in  the case of
Cir  X-1 the  ankle  appears connected  with  a sudden  change in  the
spectral state  and a related drop  in flux, as shown  by our spectral
analysis.

The flux decrease during this hard state is very rapid, being a factor
$\sim$ ten  in less  than three  days (from ObsId  04-08 to  ObsId 39,
Table~\ref{specresults}).   After  the  second periastron  passage,  a
greater mass  transfer may  have caused the  triggering of  the second
outburst.  The minor outburst evolves much more slowly with respect to
the  major  outburst.  After  ten  days,  no  significant decrease  in
luminosity is observed (between ObsId  41 and 46).  The spectral shape
is  consistent with  the  region  B parameters  of  the May  outburst.
However, at the  third periastron passage, the outburst  turns off.  A
brightening  is   observed  at  phase  3.11,   with  probably  similar
characteristics  of ObsId  03-01. The  last observations  point  to an
almost    steady     state    of    accretion     at    flux    levels
(2--3)\,$\times$\,10$^{-10}$ \ergsec.

\subsection{The mechanism behind the major outburst: 
analogies with the Be XRB systems}

After  several years  of the  monitoring, from  1996 to  2005,  of the
steady bright  X-ray state of Cir  X-1, we have found  that the source
displays   a  substantially   different  long-term   behaviour,  where
several-month-long  periods of relative  faintness are  interrupted by
bright outbursts,  whose luminosity peaks can  reach Crab-levels. This
behaviour is reminiscent of the  X-ray activity observed at the end of
seventies \citep{kaluzienski76}, for which \citet{murdin80} proposed a
model  involving  the orbital  precession  by  10$^{\circ}$/year of  a
highly  eccentric   orbit  ($e$  =   0.8\,$\pm$\,0.1).   Today's  most
widespread  interpretation of the  outbursting behaviour  of accreting
X-ray sources  is contained in the disk  instability model \citep[DIM,
][]{lasota01}. According to this model, a change in the opacity caused
by  the  ionisation  of  hydrogen  causes a  dramatic  change  in  the
viscosity of disk matter.   This instability propagates at the viscous
timescales (which  explains the  measured rise times  of the  order of
days).  When the  viscosity allows for the rapid  accretion of matter,
the outburst reaches its peak,  and following evolution (tens of days)
is driven by the emptying of  the disk, at least for the regions where
viscosity  remains  sufficiently  high.   An exponential,  or  linear,
light-curve decay  is predicted in the  case the outer  disk radius is
kept, or not, in this high viscosity regime \citep{king98,shahbaz98}.

The  brightest outbursts  in Cir  X-1 appear  to be  clocked  with the
periastron passage, and it is reasonable to argue that what drives the
instability  is a sudden  change in  the surface  mass density  at the
outer edge of  the disk.  A partial disruption of  the outer disk edge
due to a tidal interaction,  or a periastron driven enhanced accretion
\citep{regos05,  church09},  may  be  mechanisms that  can  make  this
happen, although it  would remain an open question  why this mechanism
does not  operate at every passage,  given its dependence  only on the
system particular  geometry.  The scenario  in which part of  the disk
outer mass is  stripped by a previous close  NS-companion encounter is
similarly  unable to  provide a  consistent trigger  of  the outburst,
since any  debris would probably  have already been expelled  from the
system at  the following passage, given the  high orbital periodicity.
An alternative  explanation of the episodic occurrence  is possible if
we  associate  the episodes  of  enhanced  accretion  with a  variable
structure present around  the companion star in a  scenario similar to
the classical picture  of Be X-ray binary systems  (Be XRBs). In these
systems, the accreted mass is not directly stripped by means of steady
accretion  from the  inner Lagrangian  point, but  the  compact object
accretes as it swaps the regions of a steady decretion disk around the
companion  star.  Decretion  disks are  variable in  both  density and
size,   and  they   constitute  the   only  efficient   mechanism  for
transferring  mass to  a compact  object, when  the companion  star is
neither a supergiant  with a strong wind, nor  a low-mass star filling
its Roche lobe.  Cir X-1  shares, in fact, some common characteristics
with the Be XRBs class,  such as eccentric orbit, high-mass companion,
long orbital period, enhanced  accretion at the periastron passage, as
well as  strong X-ray variability  on short and long  timescales.  The
hypothesis that  a circumstellar  decretion disk could  effectively be
present was first explored  by \citet{clarkson04}, in order to explain
the short  dip intervals that were  observed in the  X-ray light curve
scattered   around   the   zero   phase.    \citet{jonker07}   studied
spectroscopic and photometric  observation in the \textit{I}-band; the
optical spectrum  showed not only broad absorption  Paschen lines, but
also emission  Paschen lines when the  phase was near  zero.  If these
lines  are produced in  the companion  photosphere, then  a comparison
with typical  supergiant mid-B  spectral type companion  star provides
the best fit  to the optical spectrum (but  the possible spectral type
is  uncertain up  to  the A0  class).   However Be  XRBs have  earlier
spectral-type companions;  the distribution of Be  spectral classes in
Be  XRB  is  quite  strongly  peaked  around  the  B0  spectral  class
\citep{negueruela02,mcbride08}, with no examples after the B3 spectral
class.  Isolated  Be stars  do also show  later spectral types,  up to
B8/B9; it is  argued that this is not a  selection effect, but depends
on the binary  interaction of the B supergiant,  which can efficiently
lose spin angular momentum before  the onset of the X-ray active phase
\citep{zwart95}.   Most of  the Be  XRBs  NS have  high-B fields  ($>$
10$^{11}$  G), and  appear, therefore,  as  accretion-powered pulsars.
Both  of  these  properties clearly  imply  that  Cir  X-1 is  an  odd
candidate  member  of  this  class,  owing probably  to  the  physical
circumstance that Cir X-1 has  a lower magnetic field, as indicated by
the  absence of X-ray  pulsations and  presence of  bursting activity.
\citet{clarkson04} suggested  that the low  B-field could be due  to a
buried  magnetic field undergoing  a super-Eddington  accretion phase.
However,  Cir X-1  has  a  soft spectral  shape  during the  outburst,
whereas  Be XRB  are  mostly hard  X-ray  emitters, showing  typically
power-law  spectra  with  high-energy  (above 10  keV)  cut-offs.   In
addition Cir  X-1 remains  steadily very soft  during both  the rising
phase and the outburst's decline.   A likely explanation of this is in
the low B-field  of the NS, so that  accreting matter is inefficiently
channelled  into the magnetic  caps, but  uniformly spread  across the
boundary  layer via  an  accretion disk.   The  dissipation region  is
optically   thick,  resulting   in  a   thermal,  soft   spectrum,  as
demonstrated  by our  spectral analysis.  We  note that  the Be  X-ray
source,  XRB A0538-66,  which  has  a similar  orbital  period and  an
apparently high eccentricity \citep{charles83} to Cir X-1, also showed
during an outburst episode a  similar soft spectral shape and variable
local extinction \citep{ponman84}, recalling the behaviour observed in
this analysis. However, this source  is presently in a prolonged state
of  X-ray   quiescence,  hence  we  lack   more  recent  spectroscopic
observations to furtherly probe this interesting comparison.

\subsection{Physical parameters associated with the May outburst}

If  we  extrapolate   the  linear  decay  of  the   RXTE  light  curve
(Fig.\ref{flux_unabs})   of  the  outburst   at  phase   passage  0.1,
$\sim$\,1.5 days before the actual RXTE first observation, we estimate
a peak unabsorbed  bolometric flux of 9.2\,$\times$\,10$^{-9}$ \ergsec
cm$^{-2}$, corresponding  to a luminosity  of 6.7\,$\times$\,10$^{37}$
\ergsec.  Considering the 23 day  interval of the outburst, before the
change in the linear slope of  the light curve, we derive a fluence of
$\sim$\,8\,$\times$\,10$^{-3}$ erg.  The total mass accreted by the NS
during this phase is assumed to be $\sim$\,7.3\,$\times$\,10$^{23}$ g,
where the efficiency of  the accretion process $\eta$\,=\,0.1.  Linear
decays  in the  light curves  of LMXBs  outbursting sources  should be
connected only with a partial  ionisation of the inner accretion disk,
and the  outburst is  prevented from swallowing  all of the  disk. The
disk is thus  separated into two zones: a  inner \textit{hot} zone, at
high  viscosity, and  an outer  \textit{cold} zone,  at  low viscosity
\citep{king98}.  From the  calculations shown in \citet{shahbaz98}, we
can derive the physical characteristics for the \textit{hot} zone disk
\begin{equation}
M_{\rm H} (0) = \frac{2~L_p~t_{1/2}}{\eta~c^2} \quad \rm g,
\end{equation}
where $M_{\rm H}$ is the mass of the hot zone, $L_p$ is the outburst's
peak luminosity, $t_{1/2}$ is the time needed to reduce to a half the peak
luminosity, $\eta$ is the accretion efficiency and $c$ is the light velocity
\begin{equation}
\nu = 2.14  	\times 10^{-27} B_1 L_p \eta^{-1} t_{1/2}^{-1} \quad \rm cm^{-2} s^{-1},
\end{equation}
where $B_1$ is a parameter whose value is 4\,$\times$\,10$^{5}$ (cgs), $\nu$ is the kinematic
viscosity at the edge of the heating front
\begin{equation}
R_h (0) = \sqrt{\frac{B_1 L_p}{\eta c^2}}  \rm ,
\end{equation}
where $R_h (0)$ is the hot zone disk radius at time 0.

We thus derive the values

\begin{equation}
M_{\rm H} (0) = 1.4 \times 10^{24} \quad  \rm g,
\end{equation}

\begin{equation}
\nu = 4.6 \times 10^{16} \quad \rm cm^{-2} s^{-1},
\end{equation}

\begin{equation}
R_h (0) = 5.3 \times 10^{11} \quad \rm cm.
\end{equation}

The mass  of the hot  zone corresponds to  twice the value  we derived
from the  total accreted  mass needed to  power the observed  flux, so
that when Cir  X-1 enters into the hard  state, a significant fraction
of the disk mass must still be present.

\subsection{The minor outburst}

The  second  outburst,  which  was  covered only  by  Swift,  differed
significantly from  the May  outburst. It did  not start close  to the
phase-zero passage, but at phase passage $\sim$\,2.27.  The unabsorbed
fluxes associated with the spectra do not show a clear declining trend
(see Fig.\ref{flux_unabs}) and are  compatible with an average flux in
the  1--10  keV  range  of  $\sim$\,1.84  $\times$  10$^{-9}$  \ergsec
\cmmdue, whereas  in the extrapolated  0.1--100 keV range we  derive a
bolometric flux  of $\sim$\,2.1 $\times$  10$^{-9}$ \ergsec cm$^{-2}$,
corresponding to a luminosity  of 1.5 $\times$ 10$^{37}$ erg s$^{-1}$.
The outburst has  a duration of about ten days and  a total fluence of
$\sim$\,1.8 $\times$  10$^{-3}$ \ergsec, implying an  accreted mass of
$\sim$\,1.4  $\times$ 10$^{23}$  g, thus  a factor  of more  than five
lower than the  major outburst. The peak flux  is also similarly lower
than the peak flux of  the major outburst.  From the previous section,
we note that  the accreted mass during this outburst  is about 20\% of
the  total  mass of  the  \textit{hot}  zone at  the  end  of the  May
outburst.

The first snapshot  of the outburst reveals a  moderate value of local
absorption  (ObsId 41,  Table~\ref{specresults}),  possibly associated
with a passage  through a thick medium.  This  was not observed during
the first periastron passage, where accretion enhancement was strictly
confined to the  0.1 phase. It is, however,  reasonable to assume that
mass outflow  from either  the companion star  or its  decretion disk,
cause part of the ejecta to be straggled in the orbital plane where it
can be recaptured  by the NS as it  passes by \citep[see][]{lajoie11}.
The density of  these ejecta would probably be  bound with the history
of  previous  accretion phases.   Another  intertwined  effect is  the
physical conditions  in the accretion  disk.  The first passage  has a
hot  accretion disk  extending close  to  the NS  surface, whose  high
viscosity  allows matter to  rapidly accrete  when matter  is captured
within the hot  accretion region.  At the second  passage, most of the
disk is  in the cold, low  viscosity regime, and the  spreading of the
mass  raises  the  temperature  of  the  irradiated  cold  disk,  thus
triggering a new outburst.  The  decay of the outburst, with an almost
flat slope in the following days, and a very steep decline observed at
the periastron passage are, however, difficult to explain according to
the  classical disk-instability  model. The  disk returns  to  the low
accretion regime  without showing any clear trend,  at least according
to the  observational sample at hand.  In  \citet{calvelo10}, a bright
radio  flaring  was  claimed  at  this third  passage,  which  may  be
connected with  the ejection of the  inner part of the  disk, which in
turn would not allow to longer sustain the outburst.

\subsection{Comparison with previous Chandra observations}

Chandra observed the source on 4  July 2010, 11 days after the drop in
luminosity that marked the sudden end of the outburst at orbital phase
0.1.  Compared  to the set  of Swift observations of  the \textit{hard
  X-ray phase}, the spectrum no longer showed evidence for large local
absorption.  The  spectrum appears featureless, with  the exception of
probably a  narrow Gaussian line  at energies compatible  with neutral
iron K$\alpha$ emission.   The origin of this line  is probably due to
fluorescence from  the local cold  absorber. A disk  reflection origin
appears   unlikely  because   relativistically  distorted   lines  are
substantially broader, quite independently  of the luminosity state of
the accreting  source \citep{dai10}. Cir  X-1 has never been  shown to
display broad iron  lines, and a likely explanation  could be either a
high inclination angle of the system which would reduce the reflection
contribution relative  to other continuum components,  or an accretion
disk  that  does  not  extend   close  to  the  NS.   To  support  the
interpretation of a link between  the cold absorber and the iron line,
we  used  the  calculations   of  the  iron-line  equivalent-width  of
\citet{yaqoob10}.   For  a  spherically  symmetric local  absorber  of
equivalent column  density $\sim$\,3 $\times$  10$^{22}$ cm$^{-2}$, an
incident  power-law  index  of  1.46\,$\pm$\,0.11, an  iron  abundance
relative to  hydrogen of 4.67  $\times$ 10$^{-5}$, the expected  EW of
the  line is  28\,$\pm$\,1  eV,  which is  consistent  with the  value
derived from Chandra data.

The extropolated luminosity is  in the (1.26--2.11) $\times$ 10$^{36}$
erg  s$^{-1}$ range,  depending  on  the model  used  to describe  the
spectrum. This value is about a factor of two higher than that derived
from the Swift observation made  just before the onset of the outburst
(see  ObsId 40  in  Table\ref{specresults}, with  a general  agreement
between  the  spectral shape  in  the  two  observations).  A  Chandra
observation   in  January   2005  at   orbital  phase   0.10,  lasting
approximately  the same  time  as the  present  observation, when  the
source was  two times fainter,  showed emission lines from  H-like and
He-like  ions of  S,  Si,  Ar, Ca,  and  Fe \citep{iaria08}.   Another
observation in June 2005,  at phases 0.22--0.26 revealed both emission
and  absorption  features  during  turbulent flaring  episodes,  which
typically  lasted several hours  \citep{dai07a}. The  source displayed
significant  strong rapid  X-ray  variability, in  both the  continuum
shape  and  the  pattern  of  emission/absorption  lines,  covering  a
(1.6--8.5) $\times$ 10$^{-10}$ erg s$^{-1}$ cm$^{-2}$ flux range.  The
lack of  these features in  the present observation are  most probably
related to the changing  density environment through which the compact
object moves. This would also  be consistent with the continuum shape,
as in the  two 2005 Chandra observations a  partial covering component
was  needed to model  the continuum  shape, with  values in  the local
extinction  between 6.3  $\times$ 10$^{22}$  cm$^{-2}$ \citep{iaria08}
and  (8--20) $\times$  10$^{22}$ cm$^{-2}$  range  \citep{dai07a}.  We
propose  that part  of the  local  embedding neutral  matter could  be
photo-ionised near the  emitting X-ray source as observed  in the 2005
observations, while  in the present observation the  medium crossed by
the NS during this orbital passage  could be either much less dense or
far  more distant.   Interestingly,  recent radio  detections seem  to
provide further  support for this phase-dependent  effect, whose radio
compact emission from  Cir X-1 is also confined  to the 0.1--0.5 phase
range, while  the passage from  apastron to periastron  is radio-faint
\citep{moin11}.

\section{Summary}

After  a two-year long  period of  X-ray quiescence  ($L_x$ $\lesssim$
10$^{35}$ erg  s$^{-1}$), Cir X-1 went  into a bright  outburst in May
2010.   This outburst  lasted for  $\sim$\,22 days,  and  its spectral
shape  is   clearly  consistent   with  an  optically   thick  thermal
Comptonization.  During the outburst, at the periastron passage, clear
dips in the light curve and rapidly varying local absorption were both
observed.  The whole light curve  could be fitted by two linear decays
joined by an  ankle that marks at the same time  a state transition to
an  optically thin  regime,  at a  luminosity  level corresponding  to
$\sim$\,2\% $L_{\rm  Edd}$. After  this transition, the  flux decrease
became steeper, reaching a minimum flux of 0.1\% $L_{\rm Edd}$.  After
five days, a second outburst started,  with a peak flux of 7\% $L_{\rm
  Edd}$.   For the  following  ten days,  Cir  X-1 underwent  marginal
variations in flux and spectral shape.  This outburst declined steeply
in intensity  after ten  days from its  onset, following  a periastron
passage.   We  conjecture  that  a  driver of  the  outburst  sporadic
recurrence  might  have been  a  variable  decretion  disk around  the
companion  star,  by  analogy  with  the  Be  XRB  class.   A  Chandra
observation at the  end of this long outbursting  phase, at phase 0.6,
has revealed that  both the bursting activity persisted  and a lack of
photo-ionised emitting/absorbing plasma.

%%%%%%%%%%%%%%%%%%%%%%%%%%%%%%%%%%%%%%%%%%%%%%%%%%%%%%%%%
\section*{Acknowledgements}
%%%%%%%%%%%%%%%%%%%%%%%%%%%%%%%%%%%%%%%%%%%%%%%%%%%%%%%%%

Authors  express  their gratitude  to  Dr.  H.   Tananbaum for  making
possible the Chandra  DDT observations and to Dr.  N.  Schulz for kind
help and assistance
in the Chandra observation planning and set-up configuration. \\
We thank N. Gehrels and the Swift team for their
prompt response in carrying out the follow-up observations of Cir X-1.\\
This research has made use of the XRT Data Analysis Software (XRTDAS)
developed under the responsibility of the ASI Science Data Center (ASDC), Italy.\\
A.  Papitto  acknowledges the  support  of  the  operating program  of
Regione  Sardegna   (European  Social  Fund   2007-2013),  L.R.7/2007,
“Promoting  scientific   research  and  technological   innovation  in
Sardinia”, as well as of the grants AYA2009-07391 and SGR2009-811, and
of  the  Formosa  program   TW2010005  and  iLINK  program  2011-0303.
E. Egron acknowledges the support  of the Initial Training Network ITN
215212: Black Hole Universe funded by the European Community.  Authors
acknowledge  financial   contribution  from  the   agreement  ASI-INAF
I/009/10/0.  \bibliographystyle{aa} \bibliography{refs}

\newpage

\begin{table*}
\caption{Log of the RXTE observations}
\label{table_rxte_obs}
\centering      
\begin{tabular}{l c c c c c c }
\hline\hline             
ObsId                   & TSTART       & TSTOP       & Exposure & Orbital phase  & PCU2 rate    \\
                         & TJD          & TJD         & s        &                &  cts/s       \\
\hline
          95422-01-01-00         & 15327.0261   & 15327.0375  & 545      &  0.3020              &  637$\pm$14  \\
\phantom{95422-01-}01-01 & 15328.2731   & 15328.2795  & 649      &  0.3775              &  622$\pm$12  \\
\phantom{95422-01-}01-02 & 15329.4851   & 15329.5181  & 2896     &  0.4509              &  535$\pm$18  \\
\phantom{95422-01-}02-01 & 15330.0683   & 15330.0806  & 1072     &  0.4862              &  620$\pm$25  \\
\phantom{95422-01-}02-03 & 15331.4670   & 15331.4840  & 1392     &  0.5709              &  433$\pm$20  \\
\phantom{95422-01-}02-02 & 15332.1589   & 15332.1938  & 2912     &  0.6128              &  505$\pm$22  \\
\phantom{95422-01-}02-00 & 15333.5422   & 15333.5787  & 2992     &  0.6965              &  457$\pm$40  \\
\phantom{95422-01-}02-04 & 15336.0843   & 15336.1445  & 5008     &  0.8504              &  366$\pm$40 \\
\phantom{95422-01-}03-00 & 15337.1290   & 15337.1347  & 358      &  0.9136              &  331$\pm$6  \\
\phantom{95422-01-}03-01 & 15339.6255   & 15339.6466  & 1856     &  1.0648              &  194$\pm$140 \\
\phantom{95422-01-}03-02 & 15341.0607   & 15341.0701  & 736      &  1.1517              & 380$\pm$50   \\
\phantom{95422-01-}03-03 & 15343.3535   & 15343.3785  & 2112     &  1.2904              &  141$\pm$29  \\
\phantom{95422-01-}03-04 & 15343.7463   & 15343.7911  & 3760     &  1.3142              & 133$\pm$24 \\
\phantom{95422-01-}04-00 & 15344.2043   & 15344.2240  & 1600     &  1.3420              &  121$\pm$37  \\
\phantom{95422-01-}04-01 & 15344.5968   & 15344.6351  & 3280     &  1.3657              &  100$\pm$4  \\
\phantom{95422-01-}04-04 & 15345.1303   & 15345.1362  & 592      &  1.3980              & 98$\pm$3     \\
\phantom{95422-01-}04-03 & 15345.3817   & 15345.4057  & 2096     &  1.4132              & 84$\pm$3    \\
\phantom{95422-01-}04-02 & 15345.9140   & 15345.9470  & 2896     &  1.4455              &  82$\pm$3    \\
\phantom{95422-01-}04-05 & 15346.3627   & 15346.3848  & 1920     &  1.4726              & 72$\pm$3  \\
\phantom{95422-01-}04-06 & 15346.9511   & 15347.0578  & 7136     &  1.5083              & 53$\pm$3   \\
\phantom{95422-01-}04-07 & 15347.1166   & 15347.1622  & 1856     &  1.5183              & 47$\pm$2    \\ 
%change state
\phantom{95422-01-}04-08 & 15347.7357   & 15347.8418  & 7040     &  1.5558              & 24$\pm$2     \\
\phantom{95422-01-}04-13 & 15347.9998   & 15348.0368  & 3264     &  1.5717              & 21$\pm$2     \\
\phantom{95422-01-}04-10 & 15348.0961   & 15348.1030  & 592      &  1.5776              & 21$\pm$2      \\
\phantom{95422-01-}04-09 & 15348.5857   & 15348.6235  & 3248     &  1.6072              & 14.9$\pm$1.6 \\
\phantom{95422-01-}04-12 & 15349.6315   & 15349.7188  & 5296     &  1.6705              &  9.2$\pm$1.5 \\
\phantom{95422-01-}04-11 &  15350.6120  & 15350.7166  & 6768     &  1.7299              &  6.5$\pm$1.5    \\
\hline
\hline                               
\end{tabular}
\end{table*}

\begin{table*}
\caption{Log of the SWIFT/XRT observations}
\label{table_swift_obs}
\centering      
\begin{tabular}{c c c c c c c }
\hline\hline             
Obs.ID       & Mode & TSTART & TSTOP                    & Exposure & Orbital phase & Rate \\
             &         & TJD          & TJD                & s        &               &  cts/s       \\
\hline
00030268031  & WT    & 15343.824    & 15344.158          &  4498    &  1.319-1.339      & 7.12 $\pm$ 0.04  \\
00030268032  & WT    & 15344.284    & 15344.513          &  4283    &  1.347-1.361      & 4.90 $\pm$ 0.03 \\   
00030268033  & WT    & 15344.757    & 15344.982          &  3469    &  1.375-1.389      & 7.99 $\pm$ 0.05 \\   
00030268034  & PC    & 15345.709    & 15346.091          &  3602    &  1.433-1.456      & 1.41 $\pm$ 0.02 \\   
00030268038  & WT    & 15349.379    & 15349.794          &  1023    &  1.655-1.680      & 0.549 $\pm$ 0.009\\  
00030268039  & PC    & 15353.726    & 15353.866          &  2010    &  1.918-1.927      & 0.067 $\pm$ 0.006 \\ 
00030268040  & PC    & 15357.351    & 15357.480          &  3450    &  2.134-2.137      &  0.511 $\pm$ 0.011 \\ 
00030268041  & WT    & 15359.484    & 15359.623          &  3824    &  2.267-2.275      &  11.10 $\pm$ 0.01 \\ 
00030268042  & WT    & 15361.344    & 15361.564          &  1803    &  2.380-2.393      & 24.06 $\pm$ 0.2  \\  
00030268043  & WT    & 15363.488    & 15363.568          &  2162    &  2.509-2.514      & 20.55 $\pm$ 0.014 \\ 
00030268044  & WT    & 15365.371    & 15365.633          &  2033    &  2.623-2.639      & 21.95 $\pm$ 0.17 \\ 
00030268045  & WT    & 15367.300    & 15367.317          &  1466    &  2.740-2.741      & 20.90 $\pm$ 0.12 \\
00030268046  & WT    & 15369.113    & 15369.262          &  2930    &  2.850-2.859      & 18.00 $\pm$ 0.10 \\
00030268047  & PC     & 15371.452    & 15371.597          &  2828    &  2.991-3.000      & 1.05 $\pm$ 0.02 \\
00030268048  & PC     & 15373.465    & 15373.609          &  2630    &  3.113-3.122       & 0.30 $\pm$ 0.01  \\  
00030268049  & PC     & 15375.275    & 15375.345          &  1201    &  3.222-3.227       & 0.65 $\pm$ 0.02 \\ 
00030268050  & PC     & 15377.820    & 15377.898          &  1914    &  3.377-3.382       & 0.32 $\pm$ 0.01 \\
\hline
\hline                               
\end{tabular}
\end{table*}

\begin{table*}
\caption{Spectral best-fit results}
 \label{specresults}
\centering      
\begin{tabular}{l c c c c cc c c c c }
\hline\hline             
ObsId (range keV)   &  Phase & $f$\tablefootmark{a} &  $N_{\rm H}$         & $kT_{0}$  & $kT_e$   &  $\tau$    &  Fe EW         & Flux\tablefootmark{b}                                                & $\chi^2$ (dof)  \\
Units               &        &  &  10$^{22}$ cm$^{-2}$ & keV         & keV        &           &  eV             & 10$^{-10}$ erg cm$^{-2}$ s$^{-1}$ &   \\
\hline
01-00 (3-25) & 0.302 & & 3.2$\pm$1.1 & 0.84$\pm$0.07 & 2.74$\pm$0.08 & 11.7$\pm$0.5 &           & 84.3$\pm$0.5  & 1.08 (44)  \\ 
\hline
01-01 (3-25) & 0.377 & & 1.0 & 0.97$\pm$0.03        & 2.60$\pm$0.13 & 10.0$\pm$0.8 & 70$\pm$30 & 70$_{-4}^{+2}$& 0.88 (44)\\ 
01-02 (3-25) & 0.451 & & 1.0 & 0.94$\pm$0.03        & 2.53$\pm$0.07 & 11.2$\pm$0.6 & 66$\pm$30 & 61$\pm$1  & 0.65 (44)\\
02-01 (3-25) & 0.486 & & 1.0 & 1.02$\pm$0.05        & 2.60$\pm$0.09 & 12.3$\pm$0.9 & 40$\pm$40 & 72$\pm$1  & 0.50 (44)\\ 
02-03 (3-25) & 0.571 & & 1.0 & 0.92$\pm$0.05        & 2.52$\pm$0.08 & 12.4$\pm$0.8 & 60$\pm$40 & 50$\pm$1  & 1.04 (44)\\
02-02 (3-25) & 0.613 & & 1.0 & 0.98$\pm$0.05        & 2.65$\pm$0.06 & 13.3$\pm$0.6 & 40$\pm$30 & 59$\pm$1  & 0.65 (44)\\ 
02-00 (3-20) & 0.696 & & 1.0 & 0.89$\pm$0.06        & 2.59$\pm$0.05 & 15.3$\pm$0.7 & 50$\pm$30 & 54$\pm$1  & 0.76 (35)\\ 
02-04 (3-20) & 0.850 & & 1.0 & 0.81$\pm$0.07        & 2.55$\pm$0.04 & 16.4$\pm$0.7 & 50$\pm$30 & 44$\pm$1  & 0.63 (35)\\
03-00 (3-20) & 0.914 & & 1.0 & 0.81$\pm$0.07        & 2.55$\pm$0.08 & 17.3$\pm$1.3 & 60$\pm$40 & 40$\pm$0.5& 0.77 (35)\\
\hline
03-01\tablefootmark{c}            & 1.065 & &        &  &   &  &      &     & \\ 
03-02  (3-20)& 1.152 & & 5.1$\pm$0.5 & 1.0$_{-0.4}^{+0.1}$ & 2.70$\pm$0.10 & 16.0$\pm$1.1 &  & 52$\pm$1 & 0.61 (35)\\ 
\hline
03-03 (3-20)   & 1.290 & & 1.0  & 1.12$\pm$0.05 & 2.73$\pm$0.15 & 12.2$\pm$1.2 & &  16.2$\pm$0.1  & 0.80 (36)\\ 
03-04/31 (1.5-20) & 1.314 & & 2.96$\pm$0.20& 0.84$\pm$0.05 & 2.53$\pm$0.06 & 13.8$\pm$0.7 & &  18.7$\pm$0.1 [12.7\tablefootmark{d}] & 1.11 (446)\\   
04-00/32 (1.5-20)& 1.342 & & 2.98$\pm$0.25& 0.81$\pm$0.05 & 2.65$\pm$0.11 & 12.4$\pm$0.8 & &  14.7$\pm$0.1 [10.3\tablefootmark{d}]  & 0.97 (332)\\                      
04-01/33 (1.5-20)& 1.366 & & 2.18$\pm$0.18& 0.78$\pm$0.05 & 2.67$\pm$0.09 & 12.6$\pm$0.7 & &  12.6$\pm$0.1 [9.5\tablefootmark{d}]  & 1.00 (425)\\     
04-04 (3-20)   & 1.398 & & 1.0  & 0.73$\pm$0.16 & 2.28$\pm$0.12 & 17.5$\pm$2.1 & 90$\pm$70 & 11.6$\pm$0.2  & 0.74 (35) \\
04-03 (3-20)   & 1.413 & & 1.0  & 0.76$\pm$0.09 & 2.50$\pm$0.10 & 15.2$\pm$1.1 & 90$\pm$60 & 10.0$\pm$0.2  & 1.19 (35) \\
04-02/34 (1.5-20)& 1.445 & & 1.08$\pm$0.17& 0.79$\pm$0.06 & 2.65$\pm$0.10 & 12.6$\pm$0.8 & &  10.7$\pm$0.2 [10.7\tablefootmark{d}] & 0.98 (120)\\ 
04-05 (3-20)   & 1.473 & &1.0  & 0.73$\pm$0.09   & 2.35$\pm$0.09 & 16$\pm$1.2   & 80$\pm$50 & 8.6$\pm$0.2 & 0.88 (35) \\
04-06 (3-20)   & 1.508 & &1.0 & 0.80$\pm$0.06   & 2.42$\pm$0.08 & 14.6$\pm$0.9 & 90$\pm$50 & 6.2$\pm$0.1 & 1.11 (35) \\
04-07 (3-20)   & 1.518 & &1.0 & 0.81$\pm$0.13   & 2.46$\pm$0.25 & 14.3$\pm$2.3 & 140$\pm$60 & 5.4$\pm$0.2& 1.24 (35) \\
\hline %cambiamento stato
04-08 (3-20)   & 1.559 & &             &                 &               &               &            &  3.05$\pm$0.1& \\
04-13 (3-20)   & 1.572 & &             &                 &               &               &            &  2.62$\pm$0.1& \\
04-10 (3-20)   & 1.578 & &             &                 &               &               &            &  2.70$\pm$0.1& \\
04-09 (3-20)   & 1.607 & &             &                 &               &               &            &  1.93$\pm$0.1&   \\
04-12/38 (0.7-20) & 1.670 & & 1.1$_{-0.2}^{+0.3}$& 0.6$\pm$0.05 & 20        & 2.85$\pm$0.5  & 180$\pm$110&  1.12$\pm$0.1  [0.6\tablefootmark{d}] & 1.20 (136)\\
04-11 (3-20)   & 1.730 & &             &                 &               &               &            &  0.78$\pm$0.1 &   \\ 
\hline
39 (1-9) & 1.918 & 0.95$\pm$0.4           & 16$\pm$6   & 0.6                  & 20 & 2.85         &  & 0.24$\pm$0.07 & 0.72 (25) \\
40 (1-9) & 2.134 & 0.89$\pm$0.2           & 12$\pm$2   & 0.6                  & 20 & 2.85         &  & 1.3$\pm$0.2   & 0.64 (39) \\
\hline
41 (1.5-9)&2.267 & & 4.8$\pm$0.25   & 0.90$\pm$0.07  & 2.6 & 8.0$\pm$0.9   &  & 18$\pm$1  & 1.09 (533) \\
42 (1.5-9)&2.380 & &1.46$\pm$0.16   & 0.85$\pm$0.05  & 2.6 & 9.6$\pm$0.7 &  & 22$\pm$1  & 1.26 (423)     \\
43 (1.5-9)&2.509 & &1.20$\pm$0.13   & 0.90$\pm$0.05  & 2.6 & 8.7$\pm$0.8 &  & 18$\pm$1 & 1.10 (892)      \\
44 (1.5-9)&2.623 & &1.30$\pm$0.15   & 0.89$\pm$0.06  & 2.6 & 10.9$\pm$0.8 & & 20$\pm$1  & 1.45 (593)    \\ 
45 (1.5-9)&2.740 & &1.31$\pm$0.18   & 0.85$\pm$0.02  & 2.6 & 11.6$\pm$0.9 & & 19$\pm$1  & 0.97 (398)    \\ 
46 (1.5-9)&2.850 & &1.22$\pm$0.13   & 0.91$\pm$0.06  & 2.6 & 10.7$\pm$0.8 & & 18$\pm$1 & 0.99 (762)   \\
\hline
47\tablefootmark{e}  (0.7-9) &2.991 &0.82$_{-0.1}^{0.07}$ & 9$_{-3}^{+5}$  & 0.40$\pm$0.10 & 2.6 & 7.6$\pm$0.9 & 460$\pm$200 & 3.8$\pm$0.5 & 1.25 (104) \\
47\tablefootmark{f} (0.7-9)  &2.991  &0.80$_{-0.1}^{0.05}$ & 10$_{-4}^{+7}$ & 0.40$\pm$0.09 & 20    & 1.2$\pm$0.7 & 760$\pm$200 & 4$\pm$1 & 1.25 (104) \\
48\tablefootmark{e} (0.7-9)  &3.113  &0.97$_{-0.3}^{0.1}$  & 63$\pm$8 & 0.26$_{-0.10}^{0.08}$ & 2.6& 14.3$\pm$1.5 &300$\pm$150 & 8.5$\pm$2 & 1.10 (60)   \\
48\tablefootmark{f}  (0.7-9) &3.113 &0.96$_{-0.2}^{0.1}$  & 58$\pm$8 & 0.27$_{-0.10}^{0.08}$ & 20 & 4.4$_{-1.3}^{+1.5}$& 300$\pm$150 & 11$\pm$2 & 1.05 (60)   \\
49\tablefootmark{f}  (0.7-9) &3.222 & & 26$\pm$7       &$<$3 & 20 & 2.1$_{-2}^{+5}$         &    &  4$\pm$2   & 1.53 (25) \\
50\tablefootmark{f} (0.7-9)   &3.377 & & 24$\pm$7 & $<$3    & 20  &  $>$17                 &  &  2.3$\pm$0.3 & 0.80 (20) \\
\hline
Chandra\tablefootmark{f}  (0.5-9) & 3.582  &  &  2.9$_{-0.5}^{+0.7}$ &  0.81$\pm$0.16 & 20   & 6$\pm$2    & 65$\pm$55  &  2.8$\pm$0.3  &0.54 (660)  \\
\hline
\end{tabular}
\tablefoot{\tablefoottext{a}{Covering fraction of the \texttt{pcfabs} component.}
\tablefoottext{b}{Unabsorbed flux in the 2--20 keV range.}
\tablefoottext{c}{Observation analysed in Sect.\ref{sec_obs0301}.}
\tablefoottext{d}{Unabsorbed flux of  the Swift/XRT data set-}
\tablefoottext{e}{Optically thick model.}
\tablefoottext{f}{Optically thin model.}
}
\end{table*}

\end{document}